\documentclass[a4paper,oneside,11pt]{article}
\usepackage{graphicx,afterpage, comment}
\usepackage{bm}\usepackage{soul}
\usepackage{empheq}\usepackage{mathtools}
\usepackage[top=0.6in, bottom=1in, left=1in, right=1in]{geometry}
\usepackage{mathrsfs,color}
\usepackage{subcaption}
\usepackage{graphicx,afterpage}
\usepackage{bm}\usepackage{soul}
\usepackage{empheq}\usepackage{mathtools}
\usepackage[top=0.6in, bottom=1in, left=1in, right=1in]{geometry}
\usepackage{mathrsfs,color}
\usepackage{amsfonts}\usepackage{multirow}
\usepackage{amssymb}\usepackage{caption,comment}
\usepackage{amsmath}\usepackage{float}
\usepackage{amssymb,amsthm}
\usepackage{xcolor}
\usepackage{graphicx,afterpage,cancel}
\usepackage{epstopdf}
\usepackage[pdftex,colorlinks=true]{hyperref}
\usepackage{hyperref}
\usepackage{breakcites}
\usepackage[hyphenbreaks]{breakurl}
\hypersetup{
linkcolor=blue,%
citecolor=blue,%
plainpages=false,%
pdfstartview=FitH}
\usepackage{hyperref}
\hypersetup{
    colorlinks=true,
    linkcolor=blue,
    filecolor=magenta,
    urlcolor=cyan,
    pdftitle={Overleaf Example},
    pdfpagemode=FullScreen,
    }
\title{A Simple Intermediate Coupled MJO-ENSO Model: Multiscale Interactions and ENSO Complexity}
\author{Yinling Zhang, Nan Chen, and Charlotte Moser*}
\date{\today}

\begin{document}

\maketitle
\begin{center}
Department of Mathematics, University of Wisconsin-Madison\\
Corresponding author email: crmoser2@wisc.edu

\end{center}
\begin{abstract}
    The Madden-Julian Oscillation (MJO) and the El Ni\~no-Southern Oscillation (ENSO) are two dominant modes of tropical climate variability, each with profound global weather impacts. While their individual dynamics have been widely studied, their coupled interactions, particularly in the context of ENSO complexity, including spatial diversity (Central Pacific vs. Eastern Pacific events), temporal evolution (single-year and multi-year events), and intensity variations (moderate to extreme events), have received limited attention in modeling studies. In this paper, a simple intermediate coupled MJO-ENSO model is developed to address critical gaps in understanding their bidirectional feedback and its role in modulating ENSO complexity. The model integrates multiscale processes, bridging intraseasonal (MJO), interannual (ENSO), and decadal (Walker circulation) variability. Key mechanisms include: (1) interannual SST modulating MJO through latent heat and background states, (2) MJO-induced wind forcing triggering diverse ENSO events, and (3) decadal variability modulating the strength and occurrence frequency of Eastern Pacific and Central Pacific events. Effective stochastic parameterizations are incorporated to improve the characterization of multiscale MJO-ENSO interactions and the emergence of intermittency and extremes. The model captures several crucial observed MJO and ENSO features, including non-Gaussian statistics, seasonal cycles, energy spectra, and spatial event patterns. It also reproduces critical MJO-ENSO interactions: warm pool edge extension, convective activity adjustments that modulate SST, and ENSO's dependence on MJO-driven easterly and westerly wind anomalies. The model provides a useful tool to analyze long-term variations. It also advances the understanding of ENSO extreme events and their remote impacts, as well as seasonal forecasting and climate resilience.
\end{abstract}

\textbf{Plain Language Summary} The Madden-Julian Oscillation (MJO) and El Ni\~no-Southern Oscillation (ENSO) are two crucial climate patterns that affect the entire Earth system. While their individual dynamics have been studied separately, their interactions, especially in the presence of different types of El Ni\~no events (varying by location, duration, and strength) with diverse and complex features, have not received enough attention in modeling studies. This paper develops a simple intermediate coupled MJO-ENSO model that is simpler than operational models but more detailed than basic conceptual ones. The explicit spatiotemporal model structures allow the study of how MJO, ENSO, and decadal variability interact. By further incorporating appropriate stochastic components, the model successfully reproduces many observed features of both phenomena, including the ENSO diversity and complexity. It is also statistically accurate, which makes it distinct from many existing models. This new model serves as a practical tool for studying MJO-ENSO relationships under different climate conditions. The computational efficiency and statistical accuracy of this model also allow for generating long simulations, which facilitates the study of extreme events and statistical forecasting, as well as providing sufficient training data for various machine learning tasks.

\textbf{Key points}
\begin{itemize}
  \item A simple intermediate coupled model is developed to capture many observed dynamical and statistical features of ENSO complexity and MJO.
  \item By spanning decadal, interannual, and intraseasonal timescales, the model advances a detailed study of cross-scale dynamical interactions.
  \item The model provides a useful tool to analyze long-term variations, study extreme events, and offers training data for machine learning tasks.
\end{itemize}

\section{Introduction}
The El Ni\~no-Southern Oscillation (ENSO) is a dominant mode of interannual climate variability in the tropical Pacific, with far-reaching impacts on global weather patterns \cite{mcphaden2006enso, ropelewski1987global, dai2000global, mcphaden2006enso, ashok2007nino}. ENSO alternates between a warm phase, El Ni\~no, and a cool phase, La Ni\~na, which are identified by positive and negative sea surface temperature (SST) anomalies, respectively. Recent studies have called attention to ENSO complexity and diversity, which characterizes how ENSO events can vary across cycles \cite{capotondi2015understanding, timmermann2018nino}. Specifically, ENSO diversity refers to the existence of two types of ENSO events distinguished by the location of their temperature anomaly centers: Central Pacific (CP) and Eastern Pacific (EP) events \cite{ashok2007nino, kao2009contrasting, kim2012statistical}. More broadly, ENSO complexity encompasses variations in the ENSO cycle by not only spatial pattern, but also intensity and temporal evolution \cite{chen2008nino, jin2008current, barnston2012skill, hu2012analysis, zheng2014asymmetry, fang2015cloud, sohn2016strength, santoso2019dynamics}. It is essential to recognize that ENSO arises from coupled air-sea interactions; thus, both atmospheric and oceanic processes require consideration in its study \cite{bjerknes1969atmospheric}.

The Madden-Julian Oscillation (MJO) is the primary source of intraseasonal atmospheric variability in the tropics; it is a planetary-scale eastward propagating convective disturbance that plays a key role in modulating tropical rainfall and atmospheric circulation \cite{madden1971detection, madden1994observations, zhang2005madden, woolnough2019madden, zhang2020four}. As a dominant atmospheric force, MJO modulates westerly and easterly wind events over the Pacific Ocean, which affect the air-sea interactions that drive ENSO, thereby influencing the suppression or intensification of El Ni\~no and La Ni\~na events \cite{puy2016modulation, puy2019influence, tang2008mjo, mcphaden2006large, hendon2007seasonal, wang2024insights}. On the other hand, SST anomalies associated with ENSO influence convection through latent heat feedbacks, changing the intensity, duration, and spatiotemporal patterns of MJO events \cite{moon2011enso, lee2019enso}. Since these phenomena govern the Earth system on the global scale, and thus have a widespread societal and economic influence, understanding their separate dynamics as well as their interactions is crucial.

Modeling realistic ENSO events that capture the diversity and complexity of the phenomenon remains a challenge. The majority of conceptual models, intermediate coupled models (ICMs), and general circulation models (GCMs) focus on capturing EP El Ni\~no events \cite{jin1997equatorial, suarez1988delayed, battisti1989interannual, ren2013recharge, weisberg1997western, zebiak1987model, chen2017enso}. Only a few theoretical and operational models are able to realistically characterize both CP and EP El Ni\~nos \cite{geng2022enso, geng2020two, chen2022multiscale, chen2023simple, dieppois2021enso, capotondi2013enso, atwood2017characterizing}. Additionally, most existing models have difficulties in reproducing the strong non-Gaussian statistics of SST anomalies, attributed to extreme El Ni\~nos, stunting the understanding and prediction of the most impactful ENSO events \cite{jin2007ensemble,timmermann2018nino,vialard2025nino}.

Similarly, the MJO has been simulated using simple theoretical models \cite{zhang2020four} as well as more complex cloud-resolving models \cite{miura2007madden, miyakawa2014madden}. The most complex models, particularly the newest GCMs in the Coupled Model Intercomparison Project Phase 6 (CMIP6), still have difficulties in fully capturing the moisture and precipitation dynamics, the eastward propagation rate, and the intensity of MJO events \cite{lin2024assessment, le2021underestimated}. Simpler MJO models are constructed from four main theories -- skeleton theory \cite{majda2009skeleton}, moisture-mode theory \cite{adames2016mjo}, gravity wave theory \cite{yang2013triggered}, and trio-interaction theory \cite{wang2016trio}. Each theory addresses different aspects of the multiscale dynamics, the planetary-scale dry dynamics, the lower-tropospheric moisture, and the synoptic-scale convective/wave activity associated with MJO \cite{zhang2020four, jiang2020fifty}. These key features of MJO demand simulations to resolve coupled processes across disparate scales, for which stochastic parameterization is essential in capturing \cite{majda2009skeleton, thual2018tropical}.

The interactions between MJO and ENSO bridge intraseasonal, interannual, and decadal timescales \cite{hendon2007seasonal, thual2018tropical, moser2024stochastic, yang2021enso}. The complex multiscale nature of the MJO-ENSO relationship poses persistent challenges in modeling the system. Coupled GCMs have shown some success in simulating the MJO-ENSO interactions. However, biases in tropical convection and mean state representation inhibit accurate representation of the coupled system \cite{marshall2009coupled, newman2009important, capotondi2015optimal}. On the other hand, theoretical models of MJO-ENSO interactions, including both ICMs and conceptual models, have not been systematically studied, thereby limiting our understanding of the joint MJO-ENSO dynamics. The few existing theoretical models of the coupled MJO and ENSO struggle to produce realistic ENSO complexity \cite{thual2018tropical, yang2021enso}.

In this paper, we develop a simple statistically accurate ICM that captures several crucial features of ENSO complexity while providing insights into the physical mechanisms that govern MJO-ENSO interactions. This model integrates the interpretability of conceptual models with the spatial structure and physical realism that is required for practical applications. The model incorporates multiscale MJO-ENSO interactions across intraseasonal, interannual, and decadal timescales. Capturing the dynamics between and within these scales is essential in understanding how the short-term variability of MJO and the long-term fluctuations of ENSO influence each other. A vital component of this model is the two-way ocean-atmosphere feedback, which generates realistic ENSO diversity and complexity as well as physically consistent enhancement of MJO activity in the Pacific Warm Pool during El Ni\~no events. The model statistics closely match those of MJO and ENSO observations, making it a useful tool for understanding the theoretical mechanisms of the coupled MJO-ENSO system and assessing its impact on other climate phenomena. Finally, the model incorporates efficient stochastic parameterizations that ensure computational tractability for long simulations. The long time series generated by the model can thus be effectively employed to investigate the complex relationship between the MJO and ENSO for varying background states, as well as serve as a resource for machine learning-augmented climate studies.

The rest of this paper is organized as follows. Section \ref{Sec:Obs} introduces the observational data sets used for comparison with the model results, as well as the definitions adopted for classifying ENSO and wind events. Section \ref{Sec:Model} provides details about the model. The model simulation results are then analyzed and validated against observations in Section \ref{Sec:Simulation}. The paper is concluded in Section \ref{Sec:Conclusion}. Additional technical details are included in the Appendix.

\section{Observational Data and Definitions of ENSO, MJO, and Wind Events}\label{Sec:Obs}

\subsection{Observational data sets}

The oceanic data used here, including monthly SST, zonal current, and thermocline depth (calculated from potential temperature at the depth of 20$^{\circ}$C isotherm) measurements, are sourced from the GODAS dataset \cite{behringer2004evaluation} (\url{https://www.psl.noaa.gov/data/gridded/data.godas.html}). The present study solely uses oceanic anomalies, calculated by removing the monthly mean climatology of the whole analysis period. The data spans from 1982 to 2019, chosen to fall within the satellite era when observations are more reliable. For statistical analysis of ENSO events, a longer SST dataset (1950 - 2021) is used from the Extended Reconstructed Sea Surface Temperature version 5 \cite{huang2017extended} (\url{https://psl.noaa.gov/data/gridded/data.noaa.ersst.v5.html}). The spatial range of the data is confined to the tropical Pacific region ($120^{\circ}$E - $80^{\circ}$W) and is averaged meridionally within $5^{\circ}$S - $5^{\circ}$N. The eastern and western Pacific regions are defined as $120^{\circ}$E - $160^{\circ}$W and $160^{\circ}$W - $80^{\circ}$W, respectively. The Ni\~no 3 index is defined as the average SST anomaly over the region $150^{\circ}$W - $90^{\circ}$W whereas the Ni\~no 3.4 and Ni\~no 4 indices correspond to $170^{\circ}$W-$120^{\circ}$W and $160^{\circ}$E-$150^{\circ}$W, respectively.

The majority of the atmospheric data used in this study, including daily zonal wind, geopotential height, and specific humidity measurements, is obtained from the National Centers for Environmental Prediction-National Center for Atmospheric Research (NCEP-NCAR) reanalysis data set (\url{https://psl.noaa.gov/data/gridded/data.ncep.reanalysis.html}). The record extends from January 1979 to August 2024. Additionally, daily interpolated outgoing long-wave radiation (OLR) data from the National Oceanic and Atmospheric Administration (NOAA) \cite{liebmann1996description} (\url{https://psl.noaa.gov/data/gridded/data.olrcdr.interp.html}) is utilized as a surrogate for convective activity \cite{stechmann2015identifying}. The OLR observations span from June 1974 to December 2022. To ensure consistency, both datasets are analyzed over the common period from January 1, 1979, to December 31, 2022. Both datasets have a horizontal spatial resolution of $2.5^{\circ} \times 2.5^{\circ}$.

\subsection{Definitions of different types of the ENSO events}
Following the definitions in \cite{kug2009two}, El Ni\~no and La Ni\~na events are identified from the average SST anomalies during boreal winter, i.e., from December to February (DJF). Particularly, EP El Ni\~no events are characterized by EP SST anomalies above 0.5$^{\circ}$C that are larger than CP SST anomalies. Furthermore, based on the definition in \cite{wang2019historical}, an extreme EP El Ni\~no event corresponds to a maximum EP SST anomaly above 2.5$^{\circ}$C between April and the following March. Similarly, CP El Ni\~no events are defined by CP SST anomalies above 0.5$^{\circ}$C that are larger than the coinciding EP SST anomalies. On the other hand, La Ni\~na events are defined by SST anomalies below -0.5$^{\circ}$C and are similarly categorized as a CP or EP event based on the location of peak cooling. Finally, when El Ni\~no or La Ni\~na conditions occur for two or more consecutive years, it is defined as a multi-year El Ni\~no or La Ni\~na, respectively.

\subsection{Definitions of the wind events}

Wind bursts are crucial phenomena in the tropical Pacific that can influence the development and evolution of ENSO events. Wind events are detected from intraseasonal wind fluctuations, which are obtained by removing the three-month running average (interannual component) from the original wind data. Westerly wind events (WWEs) and easterly wind events (EWEs) are identified when the following criteria are met \cite{chiodi2014subseasonal,puy2016modulation}: (1) the event persists for a minimum of 3 days, (2) has a zonal extent of at least 10 degrees, and (3) has a wind magnitude exceeding a threshold of 5 m/s.

\subsection{Definitions of the MJO events}

The reconstruction of the MJO signal follows the theoretical framework developed by \cite{stechmann2015identifying}. The MJO signal is a linear combination of the zonal wind, potential temperature, convective activity, and moisture variables that are filtered to the intraseasonal planetary band. Specifically, these variables are projected to the zonal Fourier modes $1 \leq |k| \leq 3$ and the temporal frequencies $\frac{1}{90} \leq |\omega| \leq \frac{1}{30}$ cycles per day (cpd). The atmospheric states are then weighted by the MJO eigenvector for each Fourier mode $k$, denoted by $\hat e_{\text{MJO}}(k)$, which is derived through linear analysis of the MJO skeleton model \cite{majda2009skeleton} used in this work. That is, the eigenvectors indicate the contribution of each atmospheric state variable to the MJO mode. More precisely, the atmospheric state vector $\mathbf{X}(k,t)$ is projected onto the MJO eigenmode:
\begin{equation} \text{MJO}^*(k,t) = \hat e_{\text{MJO}}(k)^{\dagger} \mathbf{X}(k,t),
\end{equation}
where ${\dagger}$ denotes the conjugate transpose. Finally, an inverse Fourier transform is applied to $\text{MJO}^*(k,t)$ to obtain the MJO signal, $\text{MJO}(x,t)$, in physical space. The detailed mathematical derivations and definitions of the atmospheric state vector components are provided in the Appendix Section \ref{Appendix:Meri}.

\section{The Coupled MJO-ENSO Model}\label{Sec:Model}

\subsection{Governing equations of the coupled model}

In this subsection, we introduce the full set of equations governing the multiscale MJO-ENSO coupled model. Detailed explanations of the underlying mechanisms and dynamical processes are provided in the subsequent subsections.

The coupled system consists of the following set of stochastic partial differential equations, representing the complex interactions between the ocean and atmosphere dynamics across intraseasonal to decadal timescales:

Intraseasonal atmosphere:

\begin{equation}\label{Eq:Atm_intra_1}
\begin{aligned}
\left(\partial_t+d_u\right) u^{\prime}-y v^{\prime}-\partial_x \theta^{\prime} & =0, \\
y u^{\prime}-\partial_y \theta^{\prime} & =0, \\
\left(\partial_t+d_u\right) \theta^{\prime}-\left(\partial_x u^{\prime}+\partial_y v^{\prime}\right) & =\bar{H} a^{\prime}, \\
\left(\partial_t+d_q\right) q^{\prime}+\bar{Q}\left(\partial_x u^{\prime}+\partial_y v^{\prime}\right) & =-\bar{H} a^{\prime}+\hat\sigma_q(E_q,s_q,I)\dot{W}_q, \\
\left(\partial_t+d_a\right) a^{\prime} & =\Gamma q^{\prime}\left(\bar{a}+a^{\prime}\right) + \hat\sigma_a(\bar{a},a^{\prime},q')\dot{W}_a.
\end{aligned}
\end{equation}

Interannual atmosphere:
\begin{equation}\label{Eq:Atm_inter}
\begin{aligned}
-y \bar{v}-\partial_x \bar{\theta} & =0, \\
y \bar{u}-\partial_y \bar{\theta} & =0, \\
-\left(\partial_x \bar{u}+\partial_y \bar{v}\right) & =\delta_A\bar{H} \bar{a}-s^\theta, \\
\bar{Q}\left(\partial_x \bar{u}+\partial_y \bar{v}\right) & =-\delta_A\bar{H} \bar{a}+s^q+E_q, \\
\bar{H} \bar{a} & = \left(E_q+s^q-\bar{Q} s^\theta\right) / \left[\delta_A(1-\bar{Q})\right].
\end{aligned}
\end{equation}

Ocean:
\begin{equation}\label{Eq:Ocean_1}
\begin{aligned}
\partial_t U - c_1 Y V + c_1 \partial_x H & =c_1 \tau_x, \\
Y U+\partial_Y H & =0, \\
\partial_t H+c_1\left(\partial_x U+\partial_Y V\right) & =0.
\end{aligned}
\end{equation}

SST:
\begin{equation}\label{Eq:SST2}
\partial_t T= -c_1 \zeta E_q + c_1 \eta_1  H+\kappa(I) c_1\eta_2 U.
\end{equation}

Decadal component:
\begin{equation}\label{Eq:I}
\frac{\mathrm{d}I}{\mathrm{d} t}=-\lambda(I-m)+\sigma_I(I) \dot{W}_I.
\end{equation}

Coupling components:
\begin{equation}\label{Eq:Couplings_4}
\begin{aligned}
\tau_x & =\gamma (\bar{u}+u^{\prime}), \\
E_q & =\alpha_q T.
\end{aligned}
\end{equation}

The intraseasonal atmospheric equations \eqref{Eq:Atm_intra_1} are derived from the skeleton model for the MJO \cite{majda2009skeleton} whereas the interannual atmospheric equations \eqref{Eq:Atm_inter} follow the Matsuno-Gill model \cite{gill1980some,matsuno1966quasi}, describing the steady-state tropical atmospheric response to diabatic heating. For both atmospheric components, $u$ and $v$ represent the zonal and meridional winds, while $\theta$ is the potential temperature, $q$ is the lower-tropospheric moisture, and $a$ is the convective activity envelope. Intraseasonal atmosphere variables are denoted by a ``  $^\prime$  '' whereas interannual variables are indicated by a ``  $\bar{}$  ''. Periodic boundary conditions are used in the zonal direction. The parameter $\bar{H}$ is the convective heating rate factor, $\bar{Q}$ is a constant representing the background vertical moisture gradient, and $d_u$, $d_q$ and $d_a$ are damping parameters. The intraseasonal moisture and convective activity include stochastic components to account for unresolved atmospheric processes, where $\dot{W}_q$ and $\dot{W}_a$ are independent Gaussian white noises. The coefficients $\hat\sigma_q$ and $\hat\sigma_a$ depend on some of the state variables, and thus define state-dependent (or multiplicative) noise amplitudes. The parameter $\Gamma$ is the growth/decay rate of intraseasonal convective activity in response to fluctuations in moisture. In addition, $\delta_A$ is a dimensionless factor that scales the interannual convective heating to ensure appropriate atmospheric circulation response. Finally, $s^q$ and $s^{\theta}$ are external sources of moistening and cooling.

The ocean dynamics \eqref{Eq:Ocean_1} are governed by the shallow-water equations \cite{vallis2016geophysical} with a long-wave approximation, where $U$ and $V$ are the zonal and meridional ocean currents, while $H$ represents the thermocline depth. Reflection boundary conditions are applied at both the western and eastern boundaries (details are included in the Appendix Section \ref{Appendix:Meri}). The SST equation \eqref{Eq:SST2} extends the standard heat budget formula \cite{jin1997equatorial} by incorporating both thermocline feedback $\eta_1(x) H$ and zonal advective feedback $\eta_2(x) U$ (the explicit dependence of $\eta_1$ and $\eta_2$ on $x$ is dropped in the equations for notation simplicity).
The spatially dependent coefficient $\eta_1(x)$ peaks in the EP region because of the shallower thermocline, whereas $\eta_2(x)$ attains its maximum in the CP region due to the stronger zonal gradient of the background SST.
Notably, the zonal advection coefficient is modulated by the decadal variability $I$ through the coefficient $\kappa(I)$. The decadal component represents the background Walker circulation strength and is driven by a simple stochastic process in \eqref{Eq:I} with $\dot{W}_I$ being a standard Gaussian white noise. Equation \eqref{Eq:Couplings_4} shows the dependence of the latent heat $E_q$ on the SST and the zonal wind stress $\tau_x$ on the total wind. The time coordinate is denoted by $t$ while $x$ represents the zonal axis. The coordinate variables $y$ and $Y$ represent the meridional axes for atmospheric and ocean dynamics, respectively. Two separate meridional axes are necessary as the deformation radii for the ocean and the atmosphere are distinct, but are linked by the relation $y = \sqrt{c} Y$, where $c$ is the conversion factor between the two radii. Furthermore, the parameter $c_1 = c/\epsilon$ is a modified ratio of ocean/atmosphere phase speeds where $\epsilon$ represents the Froude number. Finally, in the coupling equations, $\zeta$ in \eqref{Eq:SST2} is the latent heating exchange coefficient and $\gamma$ in \eqref{Eq:Couplings_4} is the wind stress coefficient.

The details of these parameters and their values are listed in the Appendix Section \ref{Appendix:Meri}. 

It is worth noting that the full system \eqref{Eq:Atm_intra_1}--\eqref{Eq:Couplings_4} is formulated in two-dimensional spatial coordinates, involving both zonal ($x$) and meridional ($y$ and $Y$) directions. The system can be solved numerically by projecting the full solution onto the leading few meridional modes, which is consistent with the long-wave approximation used in the shallow water equations.
In the following discussion, we present the model as given in \eqref{Eq:Atm_intra_1}--\eqref{Eq:Couplings_4} to provide a precise physical interpretation. However, for numerical results, we project the model onto only the first meridional mode, which reduces the solution to a function of $x$ and $t$. This approximation remains appropriate for studying the leading-order behavior of MJO and ENSO, as both phenomena are dominated by zonal propagation. Such a leading-order approximation has been widely used in studying both MJO and ENSO \cite{thual2014stochastic, thual2016simple, chen2023simple, stechmann2015identifying}.

\begin{figure}[h]
    \centering
    \includegraphics[width=0.8\linewidth]{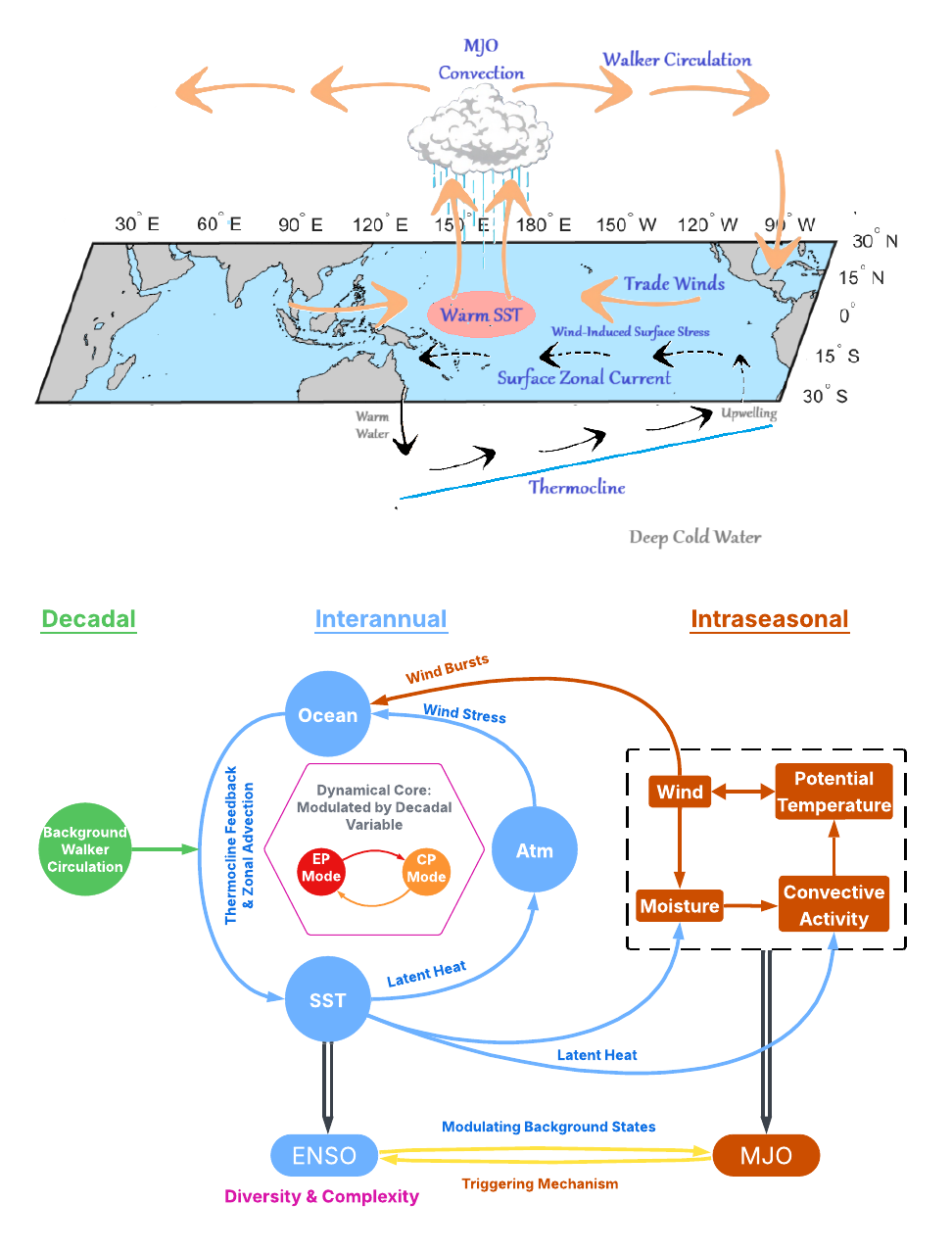}
    \caption{A schematic illustration of the multiscale model shown in \eqref{Eq:Atm_intra_1}--\eqref{Eq:Couplings_4}.}
    \label{Figure1_overview}
\end{figure}

Figure \ref{Figure1_overview} includes a schematic illustration of the coupled model.

\subsection{Interannual dynamics: The dynamical core of ENSO}\label{Sec:Dynamical_Core_ENSO}

The dynamical core on the interannual timescale provides the fundamental mechanism for generating ENSO. It involves a triad interaction between the ocean, atmosphere, and SST via \eqref{Eq:Atm_inter}--\eqref{Eq:SST2}. See the loop connecting the three blue circles in Figure \ref{Figure1_overview}. First, SST anomalies influence the atmosphere through latent heat exchange; the latent heat $E_q$ appears as a forcing in \eqref{Eq:Atm_inter} to drive the atmospheric dynamics. Second, the atmosphere affects oceanic dynamics via wind stress $\tau_x$, which appears on the right-hand side of the shallow water equation in \eqref{Eq:Ocean_1}. Finally, the oceanic dynamics modulate SST through both thermocline and zonal advective feedback in \eqref{Eq:SST2}.

The equations of the interannual atmosphere and ocean are linear. In contrast, the SST equation is weakly nonlinear due to the appearance of $I$ and a weak quadratic dependence on $T$ in $\zeta$, effectively representing cubic damping. The latter is due to fact that the relationship between SST and latent heat exchange involves complex thermodynamic processes that vary nonlinearly with temperature, particularly in regions where the simple shallow-water framework cannot fully capture the multi-layer ocean dynamics that influence surface heat fluxes \cite{zhao2021breakdown, chen2023simple}. Nevertheless, if we linearize the SST equation and choose a fixed value of $I$, we can conduct a linear analysis of the resulting system. Overall, the interannual dynamical core behaves like a damped oscillator. Depending on the choice of value for the fixed decadal variable $I$, the linear solution, which is given by the leading two eigenmodes that decay much more slowly than other modes, exhibits different spatiotemporal patterns.

\begin{figure}[H]
    \centering
    \includegraphics[width=1\linewidth]{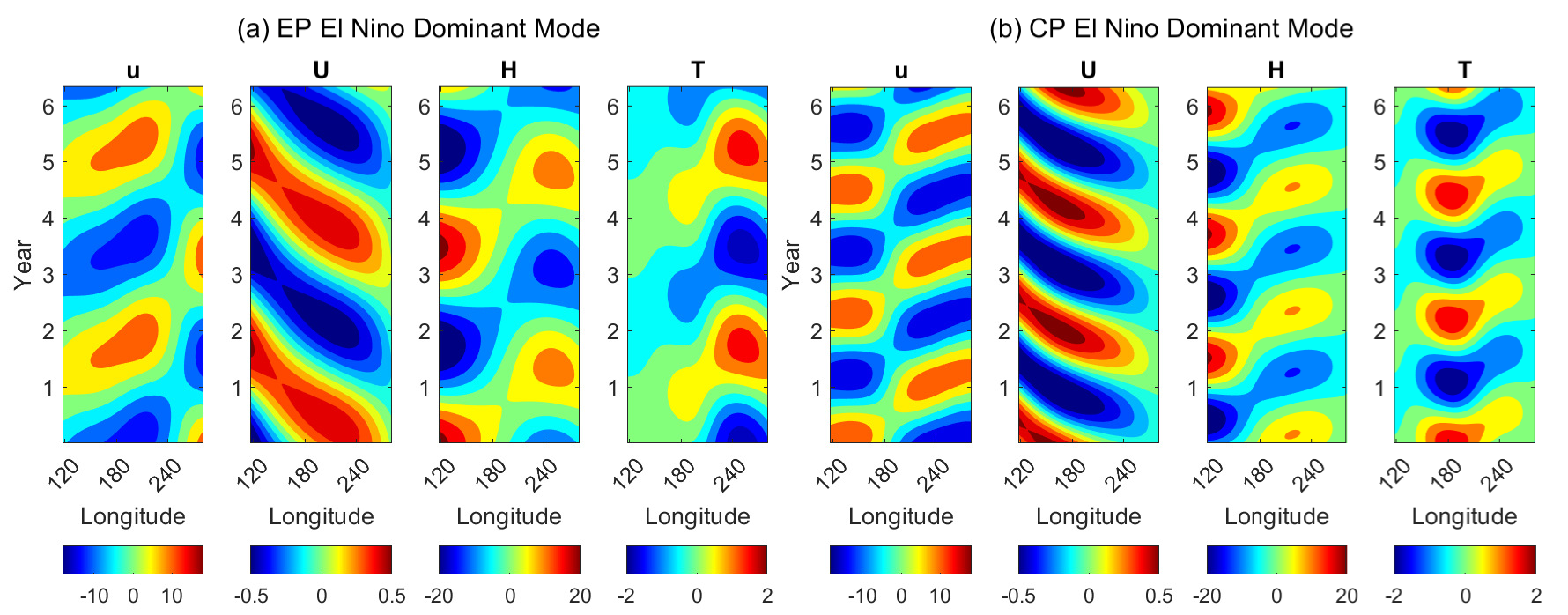}
    \caption{Linear solutions of the interannual coupled system reconstructed utilizing the leading two eigenmodes. The decay rates in these damped oscillators have been set to be zero for presentation purposes. Panel (a) shows the EP El Ni\~no dominant mode achieved by adjusting model parameters to emphasize thermocline feedback mechanisms. Panel (b) shows the CP El Ni\~no dominant mode with enhanced zonal advective feedback. In both panels, the four columns present the Hovmoller diagrams of the interannual atmospheric wind $u$, the ocean current $U$, the thermocline depth $H$, and the SST $T$. The longitude ranges from $120^{\circ}$E to $260^{\circ}$E, and the time span is approximately 6.3 years.}
    \label{fig:Figure_LinearSolution}
\end{figure}

Figure \ref{fig:Figure_LinearSolution} presents the linear solutions using  two representative values of $I$ to illustrate the fundamental behavior of the solution when $I$ is large (Panel (a)) and when I is small (Panel (b)). Specifically, Panel (a) shows the linear solution with $I=1/6$ ($\kappa(I) = 0.3$; where the decay rate in demonstrating the linear solution has been set to be zero for presentation purposes). In such a case, the zonal advective effect in the SST equation \eqref{Eq:SST2} is nearly eliminated while the thermocline feedback is the dominant component driving the SST dynamics. The resulting spatial patterns mimic the EP El Ni\~no cycles, which have a period of roughly 3.40 years. In contrast, when $I=2/3$ ($\kappa(I) = 0.9$), the zonal advection becomes more predominant. As a result, the spatiotemporal patterns shown in Panel (b) resemble CP El Ni\~no cycles, which have a period of approximately 2.13 years. These linear fundamental solutions indicate the ability of the coupled system to generate both types of ENSO events. It is anticipated that the system will trigger diverse features of ENSO when atmospheric perturbations from the intraseasonal time scale and decadal modulations are incorporated.

\subsection{Intraseasonal dynamics: The dynamical core of the MJO}

The atmospheric component of the coupled model builds upon the skeleton framework for representing MJO dynamics \cite{majda2009skeleton}. The physical variables are the zonal wind, temperature, moisture, and convective activities. In the absence of stochastic forcing terms, the only nonlinearity is in the convective activity equation through $\Gamma q' (\bar{a}+a')$. By linearizing the model and identifying the characteristic variables, four wave variables, which are linear combinations of the four physical variables, will appear. MJO is one of those characteristic variables \cite{thual2014stochastic}.

The stochastic skeleton model captures several key features of the MJO, including (a) a slow eastward propagation speed ($\sim5$ m/s), (b) a peculiar dispersion relation with near-zero group velocity ($d\omega/dk \approx 0$), (c) a quadrupole structure in the large-scale circulation, (d) the intermittent generation of MJO events, and (e) the organization of MJO events into wave trains exhibiting growth and decay.
The MJO skeleton model is derived from the three-dimensional primitive equations. Following standard practice, these equations are projected onto the first baroclinic mode in the vertical dimension \cite{thual2014stochastic, majda2009skeleton, chen2016filtering}. The stochastic skeleton model has been shown to capture the observed propagation and the statistics of the MJO \cite{stachnik2015evaluating, ogrosky2015mjo}.

The nonlinear interactions between moisture and convective activity form a nonlinear oscillator. In particular, positive moisture anomalies create a tendency to enhance the envelope of convective wave activity, while negative moisture anomalies suppress it. The convective activity envelope then feeds back into the atmospheric circulation through heating sources $\bar{H}a'$ and moisture sinks $-\bar{H}a'$, creating a closed feedback loop where enhanced convection modifies the large-scale flow. It then, in turn, affects the transport and distribution of moisture. See the bottom-right part of Figure \ref{Figure1_overview} for an illustration of this relationship. This moisture-convection feedback mechanism allows the generation of self-sustaining eastward-propagating disturbances that capture the fundamental characteristics of observed MJO events.

\subsection{Coupling from the intraseasonal to the interannual dynamics}
The intraseasonal component affects the interannual one through wind bursts, where the intraseasonal wind becomes part of the wind stress, as shown in \eqref{Eq:Couplings_4}, which in turn affects the ocean current in \eqref{Eq:Ocean_1} and, subsequently, the thermocline depth and SST. The interannual and intraseasonal effects influence ENSO dynamics in distinct ways. The interannual wind mainly helps form the ENSO regular cycles, as shown in Figure \ref{fig:Figure_LinearSolution}. In contrast, the intraseasonal wind serves as a random disturbance. In particular, westerly wind bursts (WWBs) are one of the main triggering mechanisms of EP El Ni\~no events \cite{harrison1997westerly, tziperman2007quantifying, vecchi2000tropical}, while easterly wind bursts (EWBs) have been found to terminate strong El Ni\~nos \cite{hu2016exceptionally}. These wind bursts and other convective disturbances associated with MJO activity often initiate various ENSO events or disrupt ongoing oscillations \cite{hu2019extreme}. These random disturbances transform the ocean's natural tendency from damped oscillations into sustained and irregular patterns, that reproduce the diverse spatiotemporal variability of ENSO observed in nature \cite{puy2016modulation,hendon2007seasonal,puy2019influence}.

\subsection{Coupling from the interannual to the intraseasonal dynamics}\label{Subsec:Interannual_Intraseasonal}
The interannual dynamics affect the intraseasonal components in two distinct ways.

First, the interannual convective activity $\bar{a}$ appears on the right-hand side of the intraseasonal dynamics in \eqref{Eq:Atm_intra_1}, as the moisture anomaly affects the tendency of the total convective activity \cite{thual2014stochastic}.

Next, in addition to affecting the explicit dynamics, the interannual components also influence the statistical behavior of the intraseasonal dynamics by modulating the coefficients of stochastic forcing terms, namely $\hat\sigma_q$ and $\hat\sigma_a$. On the one hand, the state-dependent noise coefficient $\hat\sigma_a(\bar{a},a^{\prime},q')= \sqrt{\lambda_a(\bar{a}+a^{\prime})|q'|}$ indicates that a stronger background state of the convective activity $\bar{a}$ will enhance the anomalies. Such a state-dependent noise was used in the MJO skeleton model \cite{thual2014stochastic}, which is crucial in triggering extreme MJO events and reproducing the observed MJO statistics \cite{stachnik2015evaluating, ogrosky2015mjo}. A rigorous mathematical justification of the noise structure can be found in \cite{chen2016filtering}. Note that, as $\bar{a}$ is influenced by $E_q=\alpha_q T$, an enhanced SST will statistically induce more variable convective activities and subsequently stronger MJO events. This is consistent with observations, as strong El Ni\~no conditions enhance moisture availability eastward from the Western Pacific Warm Pool into the Central Pacific \cite{bergman2001intraseasonal,kapur2012multiplicative}. Such a relationship connects ENSO with MJO. On the other hand, the state-dependent noise coefficient $\hat\sigma_q(E_q,s_q,I)$ is essentially a hyperbolic tangent function of the latent heat $E_q$. Since $E_q$ is proportional to SST, $\hat\sigma_q(E_q,s_q,I)$ provides a mechanism through which enhanced SST can create more moisture, which in turn affects the wind and temperature of the intraseasonal atmospheric components, strengthening MJO activity. In addition to $E_q$, the noise coefficient $\hat\sigma_q$ also depends on the source of moistening $s_q$. As $s_q$ is a function of space $x$, this dependence allows $\hat\sigma_q$ to be a spatially dependent coefficient, which helps generate more local features of the moisture anomaly. Overall, warm SST anomalies enhance surface evaporation and convective activity, while cool SST anomalies suppress convective activity \cite{zhang1995relationship,liu2020impact,moon2011enso}. In such a way, ENSO modulates the statistics of MJO \cite{dasgupta2021interannual}. Finally, the noise coefficient $\hat\sigma_q(E_q,s_q,I)$ depends on the decadal variable $I$ as well, which will be discussed in the next subsection.

\subsection{The role of decadal variability}

As is seen in Section \ref{Sec:Dynamical_Core_ENSO}, with fixed values of $I=1/6$ and $I=2/3$, the linear solution of the interannual components exhibits EP- and CP-like patterns, respectively. It has been shown that the EP and CP events were alternatively prevalent every 10–20 years over the past century \cite{yu2012identifying, dieppois2021enso}. For example, EP events were frequently observed in the 1980s and 1990s, whereas CP events occurred more consistently after 2000 \cite{chen2015strong, freund2019higher}. In addition to decadal patterns in ENSO diversity, observational studies have demonstrated shifts in other ENSO characteristics, such as event amplitude, frequency, and global teleconnection patterns \cite{an2000interdecadal,santoso2019dynamics}. It is also revealed in \cite{mcphaden2011nino, xiang2013new} that the changes in the equatorial Pacific around the 2000s, specifically a La Ni\~na-like background state with enhanced trade winds and a steeper thermocline, are favorable for the occurrence of more frequent CP El Ni\~no events. This is consistent with the findings in \cite{capotondi2015optimal}, which highlighted the importance of a La Ni\~na-like background state in CP event development based on results from a linear inverse model. The role of decadal variability in affecting the equatorial Pacific has also been emphasized in \cite{power2021decadal}.

Therefore, it is natural to include the time evolution of decadal variability in the coupled system, allowing the system to alternate between these two regimes and generate diverse features of the ENSO. Since the primary goal is not to develop a refined model for decadal variability but instead to study the influence of decadal variability on interannual and intraseasonal variables, a simple stochastic process representing the decadal signal of $I$ in \eqref{Eq:I} is incorporated into the coupled system. The stochastic process is designed so that the resulting time series of the decadal variable fluctuates between $I=0$ and $I=1$ \cite{chen2023stochastic}. The value of $I$ in the model is non-negative because the trade winds in the lower level of the Walker circulation are easterly on the decadal time scale, which means the sign of $I$ should remain the same throughout time. The damping time is set to be $5$ years, representing the decadal timescale. Note that $I$ also represents the zonal SST gradient between the Western Pacific and CP regions that directly determines the strength of the zonal advective feedback. It is the primary interaction between decadal and interannual variabilities in the coupled system.

Finally, decadal variability also modulates the statistics of the wind through its influence on the noise coefficient $\hat\sigma_q(E_q,s_q,I)$. In particular, the noise coefficient $\hat\sigma_q(E_q,s_q,I)$ strengthens when the value of $I$ is small, corresponding to more active winds which encourage strong El Ni\~no conditions in the EP.

\subsection{Seasonal phase locking}

Seasonal phase locking represents one of the crucial features of ENSO. It is characterized by the tendency for El Ni\~no and La Ni\~na events to reach peak intensity during boreal winter \cite{stein2010seasonal,tziperman1997mechanisms}. The seasonal synchronization is implemented in the coupled model through the time-dependent modulation of the latent heating coefficient $\zeta$, which serves as the damping in the SST equation \cite{thual2017seasonal}. The seasonal modulation reflects the annual migration of the Intertropical Convergence Zone, which systematically alters upwelling strength and horizontal advection processes that govern SST anomaly evolution throughout the year \cite{mitchell1992annual, stein2010seasonal}. In addition, seasonal variations in convective activity and cloud cover, resulting from background SST changes, produce damping effects through shortwave cloud radiative feedback, which can further influence ENSO intensity. These physical processes are parameterized through a set of sinusoidal components, which have a period of 1 year, naturally describing the seasonal cycle. The seasonal phase-locking also influences the atmospheric component by modulating the background conditions for intraseasonal variability \cite{hendon2007seasonal, seiki2007westerly}. It allows the model to capture the observed seasonal characteristics of westerly and easterly wind bursts in the equatorial Pacific.

\section{Model Simulation Results}\label{Sec:Simulation}

\subsection{Coupled model simulation}

\begin{figure}[H]
    \centering
    \includegraphics[width=1\linewidth]{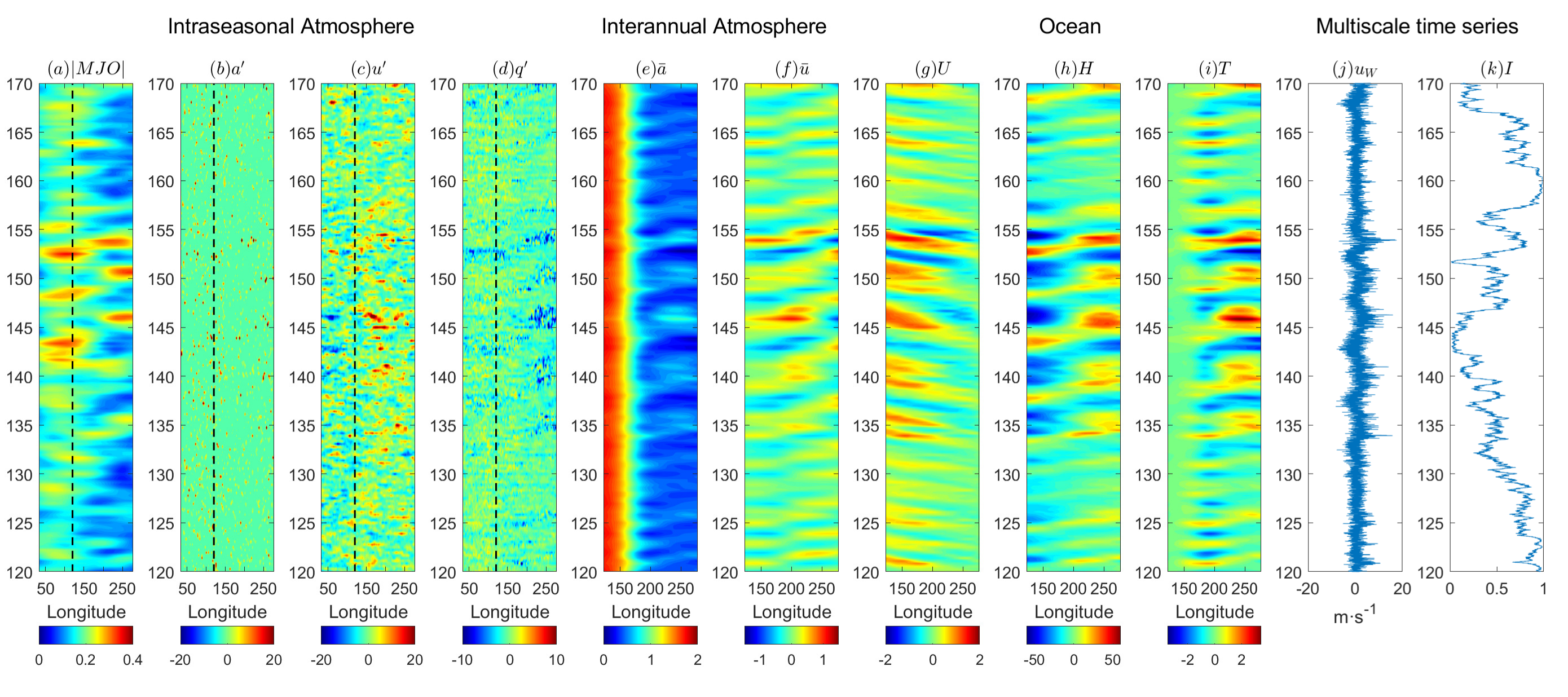}
    \caption{Hovmoller diagrams and time series showing the multiscale dynamics of the MJO-ENSO coupled model over a 50-year period (years 120-170). Panels (a)--(d) display the intraseasonal atmospheric variables: (a) a 1-year running mean of the magnitude of the MJO signal $|MJO|$; (b) intraseasonal wind $u^{\prime}$; (c) intraseasonal moisture $q^{\prime}$; and (d) intraseasonal convective activity $a^{\prime}$. Panels (e) and (f) show variables from the interannual atmosphere: (e) interannual convective activity $\bar{a}$; and (f) interannual wind $\bar{u}$. Panels (g)--(i) depict the oceanic state variables: (g) zonal ocean current $U$; (h) thermocline depth $H$; and (i) sea surface temperature $T$. Panels (j) and (k) present time series of (j) the total wind averaged over the Western Pacific $u_W$, and (k) the decadal variability $I$. In Panels (a)--(i), the x-axis denotes longitude (30$^{\circ}$E-80$^{\circ}$W for intraseasonal components, and 120$^{\circ}$E-80$^{\circ}$W for interannual components), where the black dashed line represents the Western Pacific boundary, and the y-axis shows time in years.}
    \label{Figure1_hov}
\end{figure}

Figure \ref{Figure1_hov} depicts the Hovmoller diagrams and time series of the key oceanic and atmospheric processes from the model simulation. Notably, the model successfully reproduces the observed complexity of ENSO events, as validated by the SST patterns in panel (i). Specifically, the model can generate realistic EP El Ni\~no events, both moderate (model year 140, observation year 1983) and extreme (model year 146, observation year 1998), as well as CP El Ni\~no events (model year 164, observation year 2010), mixed CP-EP events (model year 157, observation year 2006), and La Ni\~na events (model year 153, observation year 2011). Furthermore, the model can capture multi-year EP El Ni\~nos (model years 150-152, observation years 1986-1988), CP El Ni\~nos (model years 128-129, observation years 1993-1994), and La Ni\~nas (model years 132-133, observation years 1999-2000). While not shown in this frame, the model can also replicate delayed super El Ni\~nos similar to that of 2014-2015, which is shown in detail in Figure \ref{Wind_hov_case2}. It is important to note that the decadal variable (panel (k)) is essential in achieving a realistic representation of ENSO diversity. When Walker circulation is strong, such as the case in model years 157 to 166, zonal advection in the CP is enhanced, inducing CP warming. In contrast, during the simulation years 133-146, $I$ is small, zonal advection is weakened, and wind effects on SST are enhanced, encouraging the development of EP events.

The coupled model also effectively captures the complex dynamics of MJO. The large-scale patterns of the MJO amplitude (panel (a); a one-year running average) are most closely linked with the intraseasonal wind $u'$ (panel (c)), which shows increased variability in areas with a strong MJO signal. Notably, the convective activity (panel (b)) and moisture (panel(d)) are less clearly connected to MJO on broad scales, but contribute more to MJO variability on the finer scales. This is reflected by the power spectra discussed later in Figure \ref{Figure3_MJO_spectrum}, and is consistent with observations \cite{wheeler1999convectively}.

Since MJO is intrinsically linked to intraseasonal wind, the wind stress forcing in the ocean equations provides a natural pathway for MJO to influence ENSO. The eastward extension of MJO amplitude correlates with the occurrence of El Niño events, which establishes an initial connection between these two phenomena. Notably, strong MJO signals (panel (a)) are often accompanied by strong westerlies in the Western Pacific (panels (c) and (j)), which then trigger El Ni\~no events (panel (i)) (e.g., model year 146). In contrast, when MJO activity is subdued, the intraseasonal wind weakens, and La Ni\~na events are more likely (e.g., model year 158).

The influence of ENSO on MJO through latent heat completes the feedback loop between the ocean and the atmosphere. It not only correlates strong MJO activity with Pacific temperature anomalies but also enables the model to generate the eastward shift of the MJO signal during El Ni\~no events (e.g., model year 146). Furthermore, the careful use of state-dependent noises in the intraseasonal atmosphere equations establishes a clear, but non-deterministic link between the two phenomena, allowing for some years with El Ni\~no events that do not coincide with strong MJO activity (model year 137), and others with La Ni\~na events that occur while the MJO signal is strong (model year 153).

\begin{figure}[H]
    \centering
    \includegraphics[width=1\linewidth]{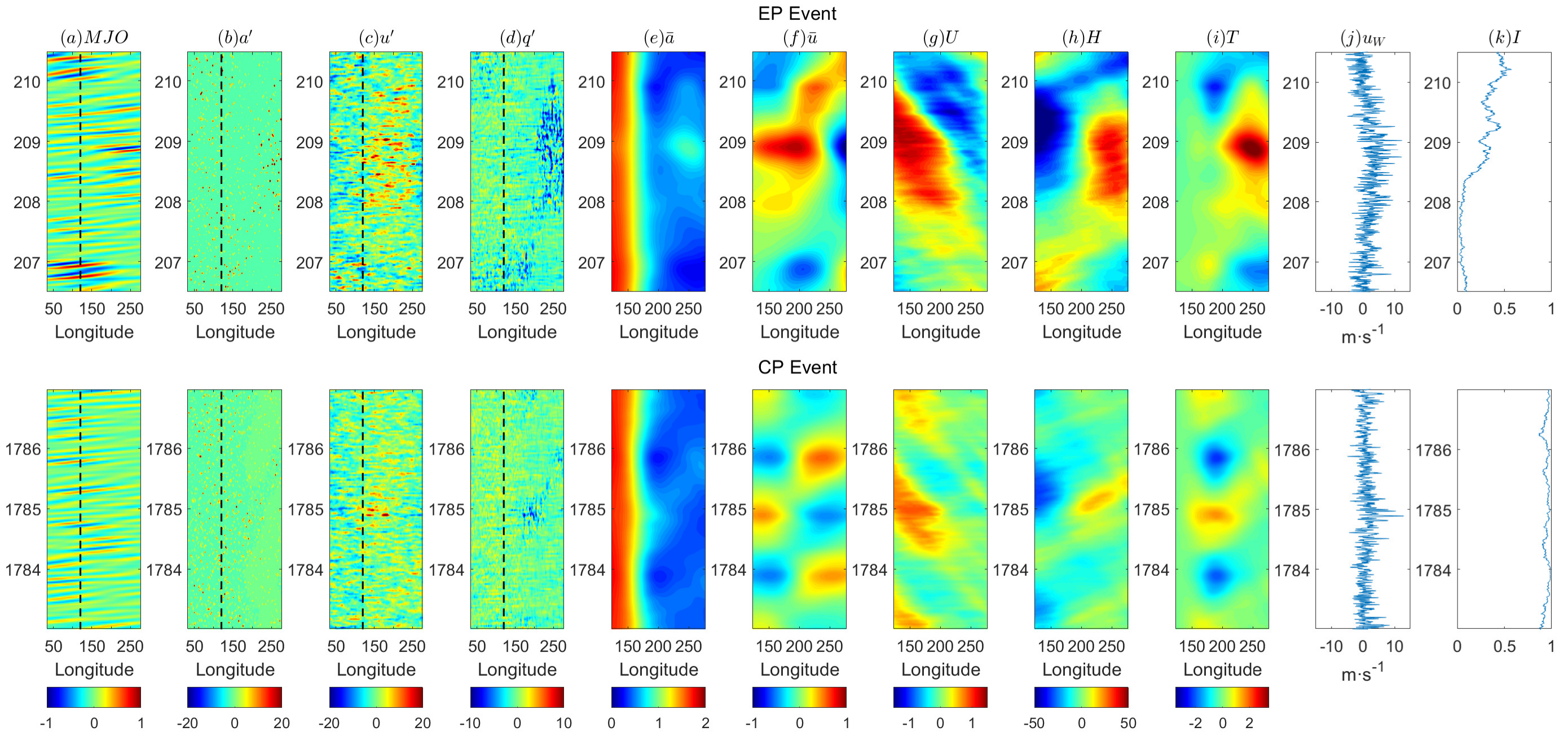}
    \caption{Model case studies of the atmospheric and oceanic variables during an EP event (top row) and a CP event (bottom row). The set up is similar to Figure \ref{Figure1_hov}, but each panel shows a 4 year period, and panel (a) shows the MJO signal. }
    \label{Figure1_hov_case}
\end{figure}

Figure \ref{Figure1_hov_case} shows the details of an EP event (top row) and a CP event (bottom row) generated by the model. Prior to the peak of both events, the MJO signal intensifies in the Pacific as a result of sea surface warming, which drives the latent heat forcing in the atmospheric equations. Furthermore, convective activity (panel (b)) drives strong westerlies in the Western Pacific leading up to the EP event in year 209 (top row in panels (c) and (j)). In contrast, strong convective activity and westerly winds coincide with the peak of the CP event, and warming is much weaker, since wind stress is less influential in the development of CP events. Such dynamical differences between CP and EP events are implemented through the decadal variable (panel (k)). In addition to modulating zonal advection, which controls CP warming, the background Walker circulation also governs the strength of the wind forcing, which contributes to the intensity of the event. When $I$ is small (large), as for the top (bottom) panel, thermocline feedback (zonal advection feedback) dominates, encouraging warming in the EP (CP). Overall, the decadal variable is strategically incorporated to produce events with different warming centers, as well as varying intensities, allowing EP events to be more severe.

\subsection{Statistics}

\begin{figure}[H]
    \centering
    \includegraphics[width=1\linewidth]{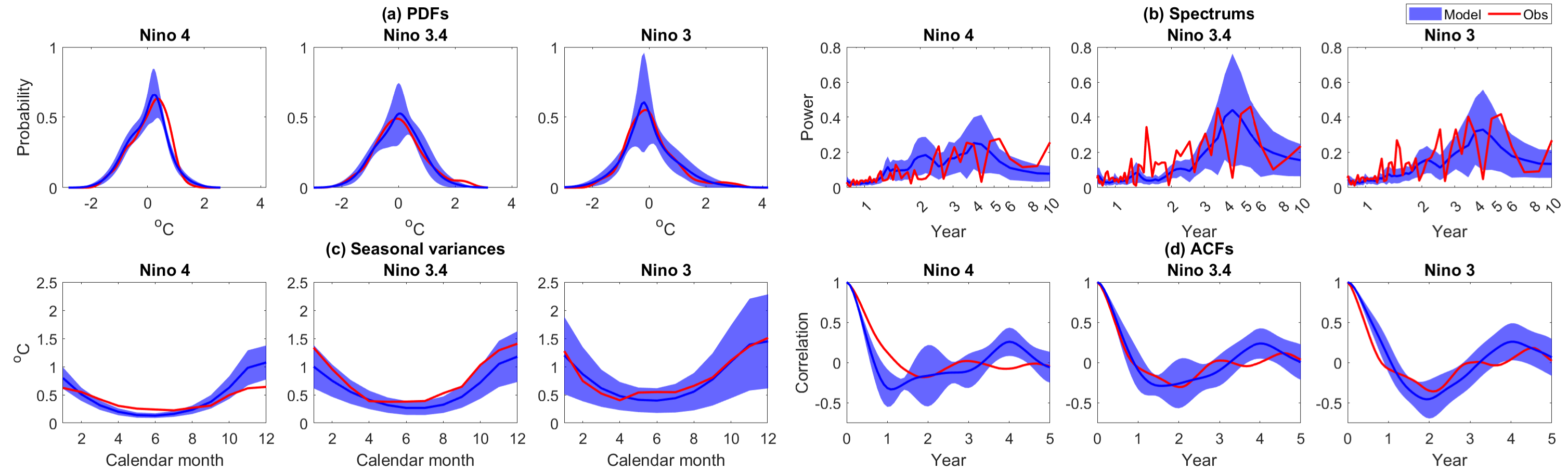}
    \caption{Comparison of statistical properties between model simulations (blue) and observations (red) for SST anomalies in the Ni\~no 4, Ni\~no 3.4, and Ni\~no 3 regions. The blue shading represents one standard deviation calculated from 50 segments of 40-year model simulations (totaling 2000 years). The observational data spans from 1980 to 2020. Panel (a) shows the probability density functions (PDFs), panel (b) the power spectra, panel (c) the seasonal variance, and panel (d) the autocorrelation functions (ACFs), i.e., the memory of the system.}
    \label{Figure2_stats}
\end{figure}

Figure \ref{Figure2_stats} illustrates the model's ability to reproduce the key statistical features of SSTs across three principal regions of the equatorial Pacific: Ni\~no 4, Ni\~no 3.4, and Ni\~no 3. Specifically, panel (a) presents the SST probability density functions (PDFs) from observation and the model simulation, showing that the model captures the negative skewness of the Ni\~no 4 distribution, the slight positive fat tail of the Ni\~no 3.4 distribution, and the large positive fat tail of the Ni\~no 3 distribution. These non-Gaussian statistics are critical in generating realistic ENSO diversity and complexity. Notably, the large positive fat tail in the Ni\~no 3 region reflects the model's success in reproducing extreme event frequencies, as shown in Table \ref{Table:Events_6d}. Furthermore, although multi-year El Ni\~no events are slightly overrepresented, the simulated event frequencies are overall highly consistent with observations. Additionally, the power spectra of all three regions (panel (b)) show dominant frequencies between 2 and 7 years, which closely match the observations. Consistent with this, the SST auto-correlation functions (ACFs) (panel (d)) illustrate the model's ability to reproduce the quasi-periodic behavior of ENSO. Another important characteristic reproduced by the model is the seasonal phase locking of ENSO, as evidenced by the peak variability of SSTs in boreal winter (panel (c)).

\begin{table}[H]
    \centering
\begin{small}
\begin{tabular}{ccccccccc}
        \hline & El Nino & EP & CP & Extreme & Multi-year & La Nina & Multi-year\\
        \hline Observation & 24 & 14 & 10 & 4 & 5 & 24 & 8 \\
        Model &  $35.0 \pm 5.0$ & $18.6\pm 6.8$ & $16.4\pm3.0$ & $3.4\pm2.7$ & $10.7\pm4.2$ & $25.4\pm3.7$ & $6.3\pm2.5$\\
        \hline
\end{tabular}
\end{small}
\caption{Occurrence frequency of different ENSO events per 71-year period. Event counts from 1950 to 2021 were collected from observational data, while the mean values and one standard deviation across 27 simulated 71-year segments are presented from the model. The table presents counts for total El Ni\~no, EP El Ni\~no, CP El Ni\~no, extreme El Ni\~no, multi-year El Ni\~no, total La Ni\~na, and multi-year La Ni\~na events.}
\label{Table:Events_6d}
\end{table}

\begin{figure}[H]
    \centering
    \includegraphics[width=1\linewidth]{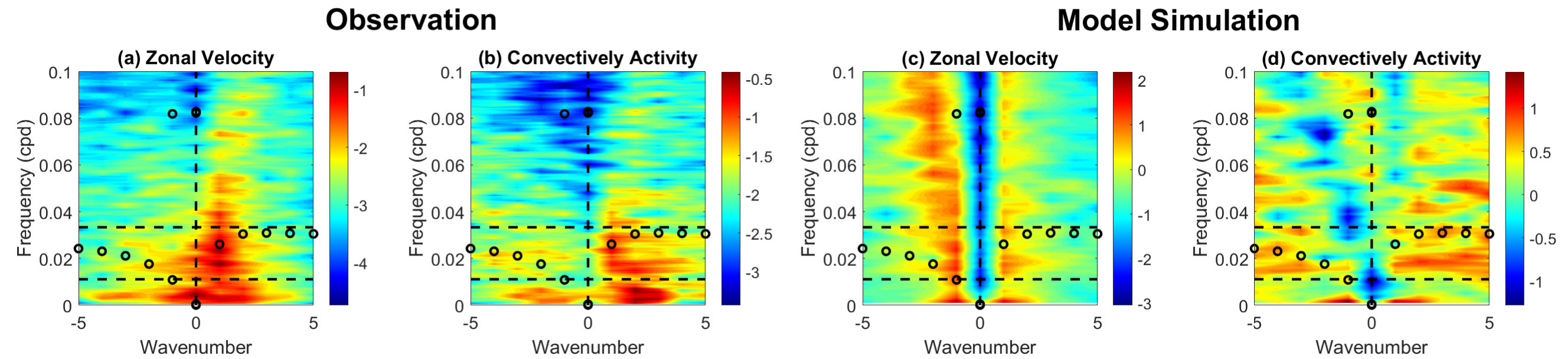}
    \caption{ Comparison of the observational (panels (a) and (b)) and model (panels (c) and (d)) power spectra for two key atmospheric variables. Panels (a) and (c) show the power spectra of the intraseasonal zonal velocity $u^{\prime}$ while panels (b) and (d) display that of the intraseasonal convective activity $a^{\prime}$. The x-axis corresponds to wavenumbers and the y-axis shows frequency in cycles per day (cpd). The two dashed horizontal lines mark the boundaries of the intraseasonal band, spanning from 30 days (0.333 cpd) to 90 days (0.111 cpd). The black dots indicate the dispersion curves from linear analysis of the MJO skeleton model. }
    \label{Figure3_MJO_spectrum}
\end{figure}

Figure \ref{Figure3_MJO_spectrum} depicts the power spectra of the intraseasonal wind and convective activity, which are the two key variables contributing to the MJO signal. For both variables, the model captures the sharp power peak associated with MJO in the intraseasonal band (1/90 to 1/30 cpd), which is marked by the dashed black lines. Within the intraseasonal band of both variables, the power is highest for the lower wave numbers $1 \leq |k| \leq 3$, reflecting the characteristic 5 m/s propagation rate of MJO. Furthermore, aligning with observations, the zonal velocity power is more concentrated at wave number one in this band, indicating that the intraseasonal wind contributes more to the large-scale spatial patterns of the MJO. In contrast, the convective activity power peak extends further beyond the first wave number, suggesting additional contributions to the smaller-scale dynamics of the MJO, consistent with observations and the qualitative assessment in Figure \ref{Figure1_hov}.

\begin{figure}[H]
    \centering
    \includegraphics[width=1\linewidth]{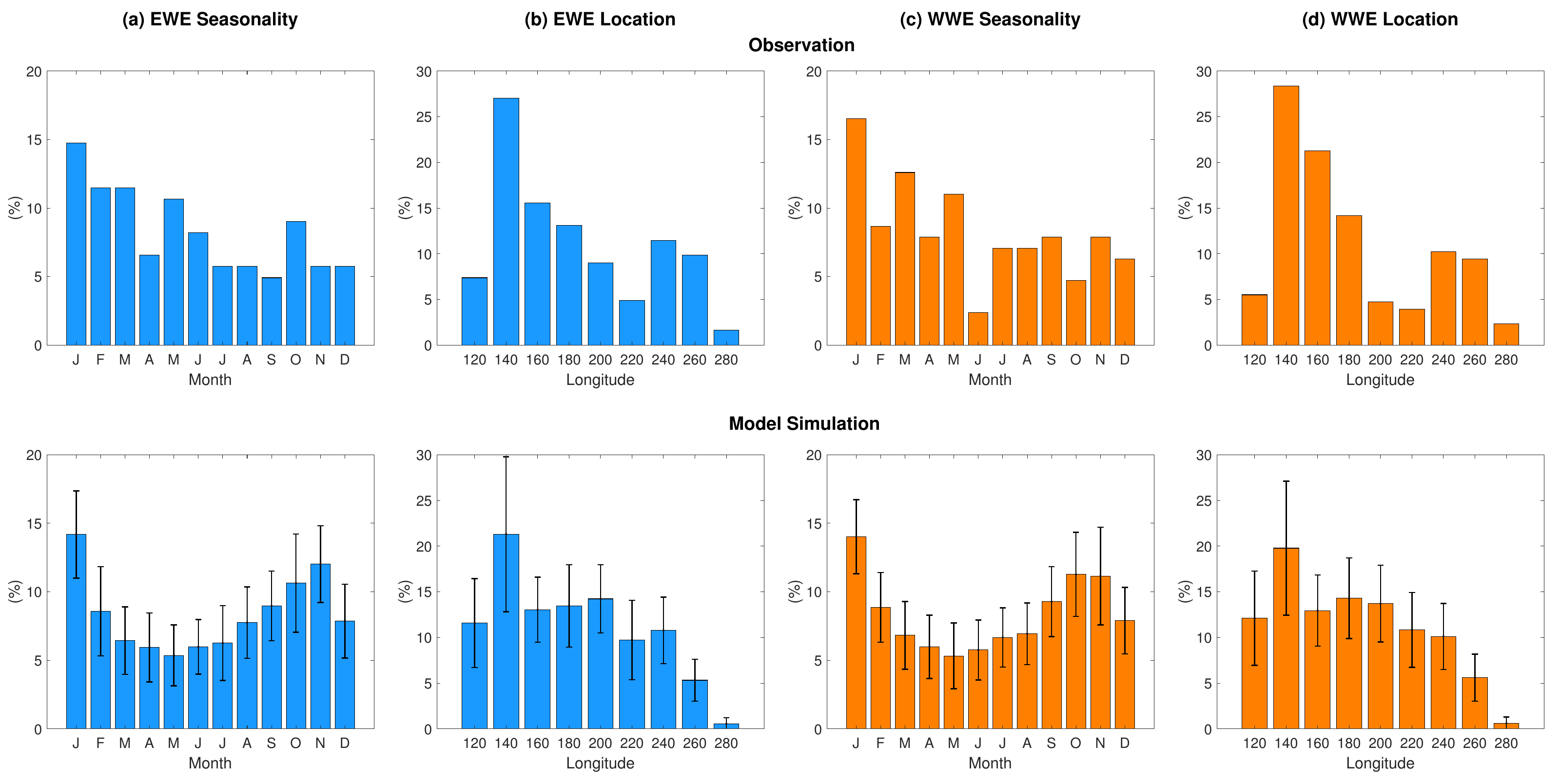}
    \caption{Seasonal and longitudinal distributions of westerly wind events (WWEs) and easterly wind events (EWEs) in observations and model simulations. The top row shows observational data from reanalysis datasets, while the bottom row displays model simulation results from 45 cycles of 44-year simulations. Panels (a) and (c) present seasonal distributions showing the percentage occurrence of EWEs (blue) and WWEs (orange) by month. Panels (b) and (d) show longitudinal distributions across the tropical Pacific (120$^{\circ}$E-280$^{\circ}$E). Error bars in model simulation panels mark one standard deviation.}
    \label{Wind_stats}
\end{figure}

\begin{figure}[H]
    \centering
    \includegraphics[width=1\linewidth]{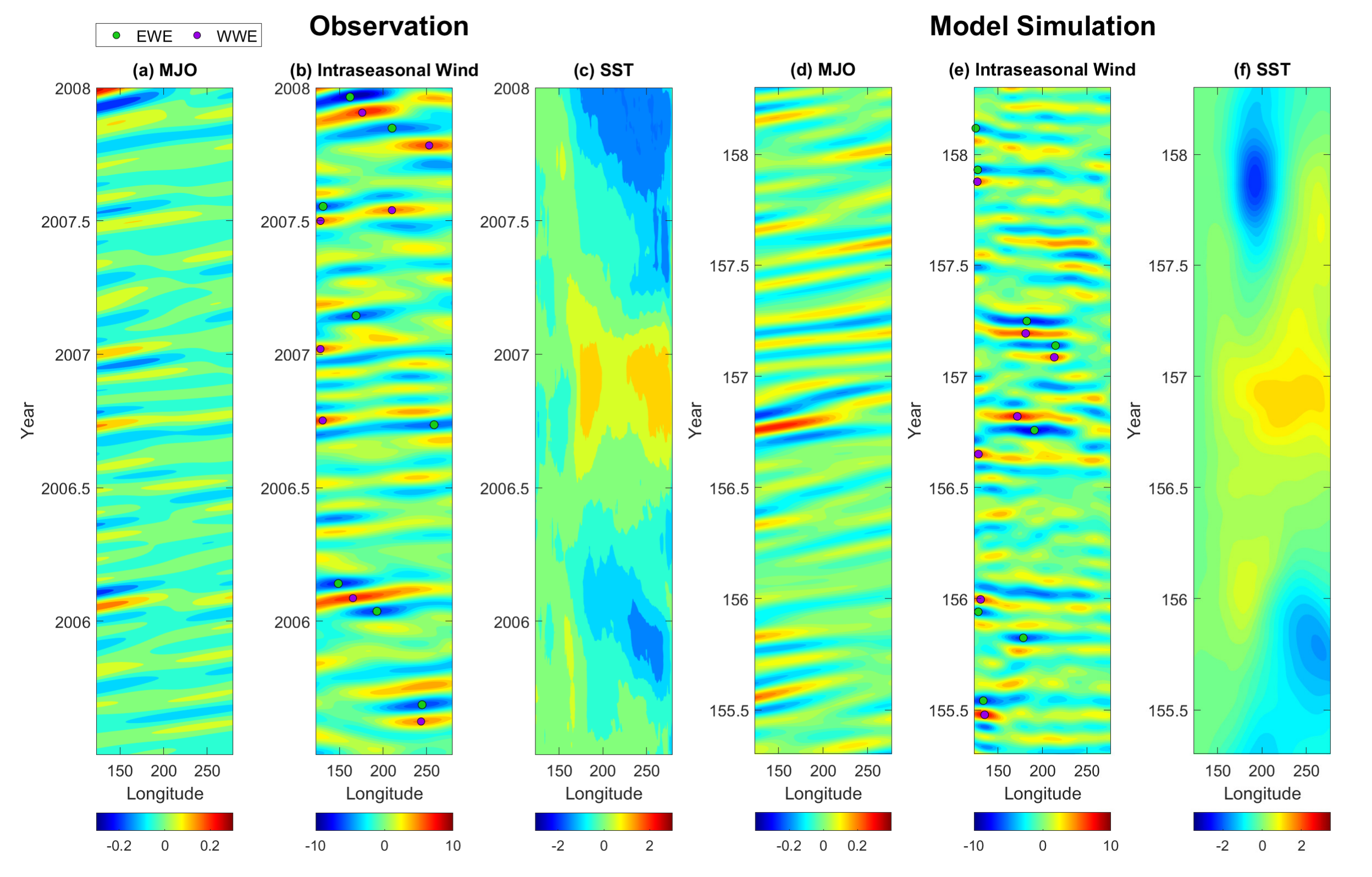}
    \caption{Spatiotemporal dynamics of the intraseasonal wind, SST, and MJO from observations (panels (a)--(c)) and the model simulation (panels (d) - (f)) for mixed CP-EP El Ni\~no events. Panels (a) and (d) show the MJO signal; panels (b) and (e) the intraseasonal winds; and panels (c) and (f) the SST anomalies. Green and red dots on the intraseasonal wind panels indicate EWE and WWE events, respectively. The x-axis marks longitude, which covers the Pacific, while the y-axis marks the year. The model MJO amplitude was scaled by a factor of $1/3$ for a cleaner visual comparison.
}
    \label{Wind_hov_case1}
\end{figure}

\begin{figure}[H]
    \centering
    \includegraphics[width=1\linewidth]{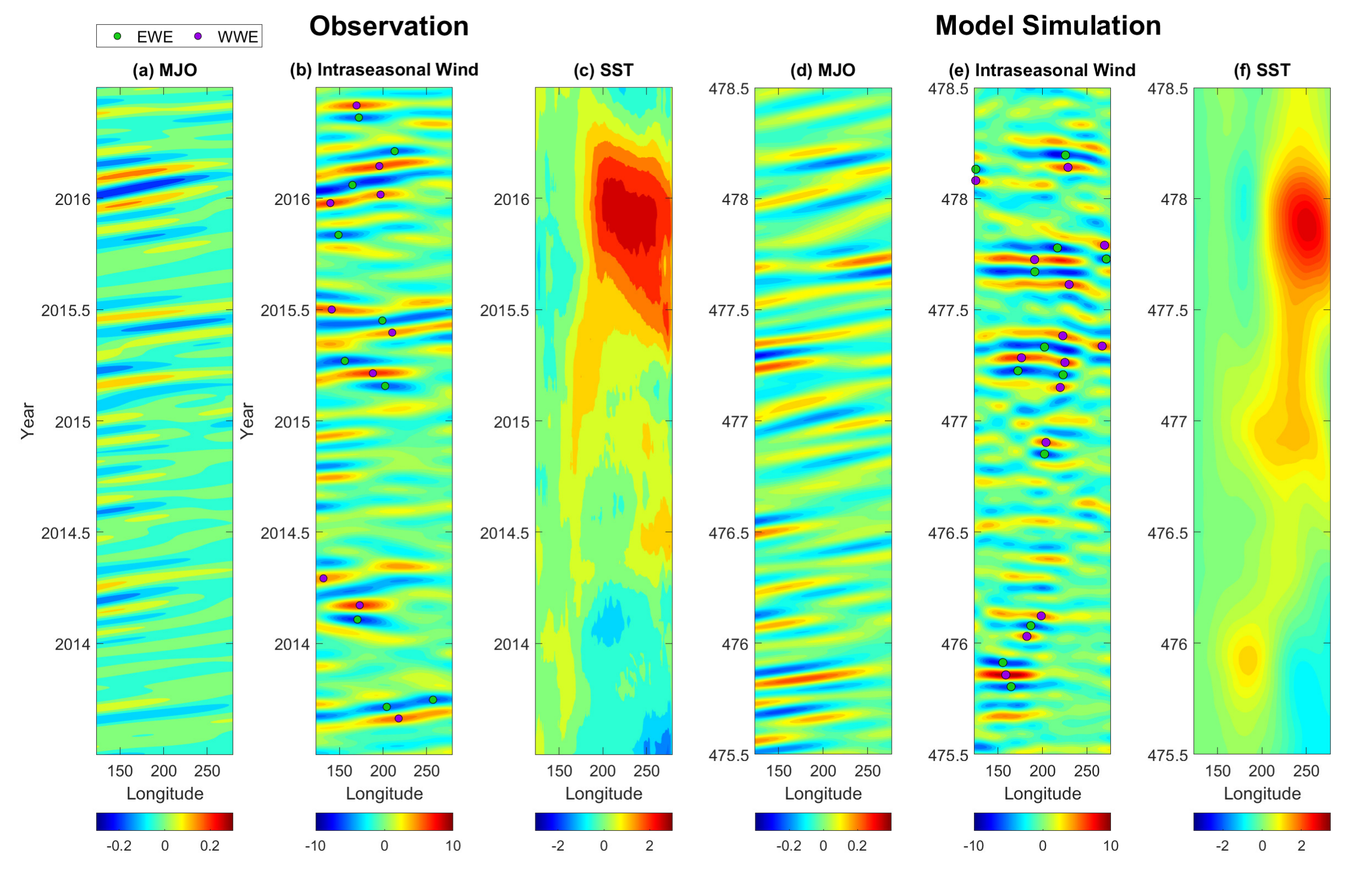}
    \caption{Similar to Figure \ref{Wind_hov_case1}, but for delayed extreme El Ni\~no events.}
    \label{Wind_hov_case2}
\end{figure}

Figures \ref{Wind_stats}--\ref{Wind_hov_case2} examine the observed and modeled wind dynamics through wind event statistics and the coupled MJO-ENSO spatiotemporal patterns. Figure \ref{Wind_stats} depicts the temporal and spatial distributions of EWE and WWE in the equatorial Pacific (see Section 2.3 for wind event definitions).  The model accurately represents the seasonality of both EWEs and WWEs, with most events occurring during boreal winter, when peak MJO and ENSO activity is present. Furthermore, the model reproduces the spatial dynamics of the wind, with events happening across the Pacific, but at higher concentrations in the Western Pacific. This demonstrates that the model accurately simulates the warm pool dynamics, where warmer SSTs excite atmospheric variability through latent heating.

Figures \ref{Wind_hov_case1} and \ref{Wind_hov_case2} depict the spatiotemporal patterns of MJO, intraseasonal wind, and SSTs during a mixed CP-EP El Ni\~no and a delayed super El Ni\~no. The intraseasonal winds for both model simulations and observations have been processed using a two-dimensional spatiotemporal Fourier transform filter, which preserves temporal oscillations in the intraseasonal band (periods between 30-90 days) while removing small-scale noise produced by high spatial wave numbers.

At the onset of the mixed events (Figure \ref{Wind_hov_case1}), both observations and model simulation show enhanced MJO activity in the Western Pacific and a larger proportion of WWEs compared to EWEs. The majority of wind events occur in the western and central Pacific. However, during the El Ni\~no event, the MJO signal extends further eastward, causing strong winds to shift further eastward, consistent with observations.

The model's air-sea interactions also allow it to produce delayed super El Ni\~nos. The delayed extreme events in both observations and the model simulation (Figure \ref{Wind_hov_case2}) exhibit a three-phase structure. First, strong MJO activity in the Pacific triggers WWEs that encourage sea surface warming. Second, a series of EWEs stunt event growth, restricting warming to the CP. Third, consecutive WWEs push the warm CP water into the EP, generating a follow-up extreme El Ni\~no event. This demonstrates that the model can represent the intricate relationship between the atmosphere and ocean that is essential in producing ENSO complexity.

\subsection{Air-sea interactions}

\begin{figure}[H]
    \centering
    \includegraphics[width=1\linewidth]{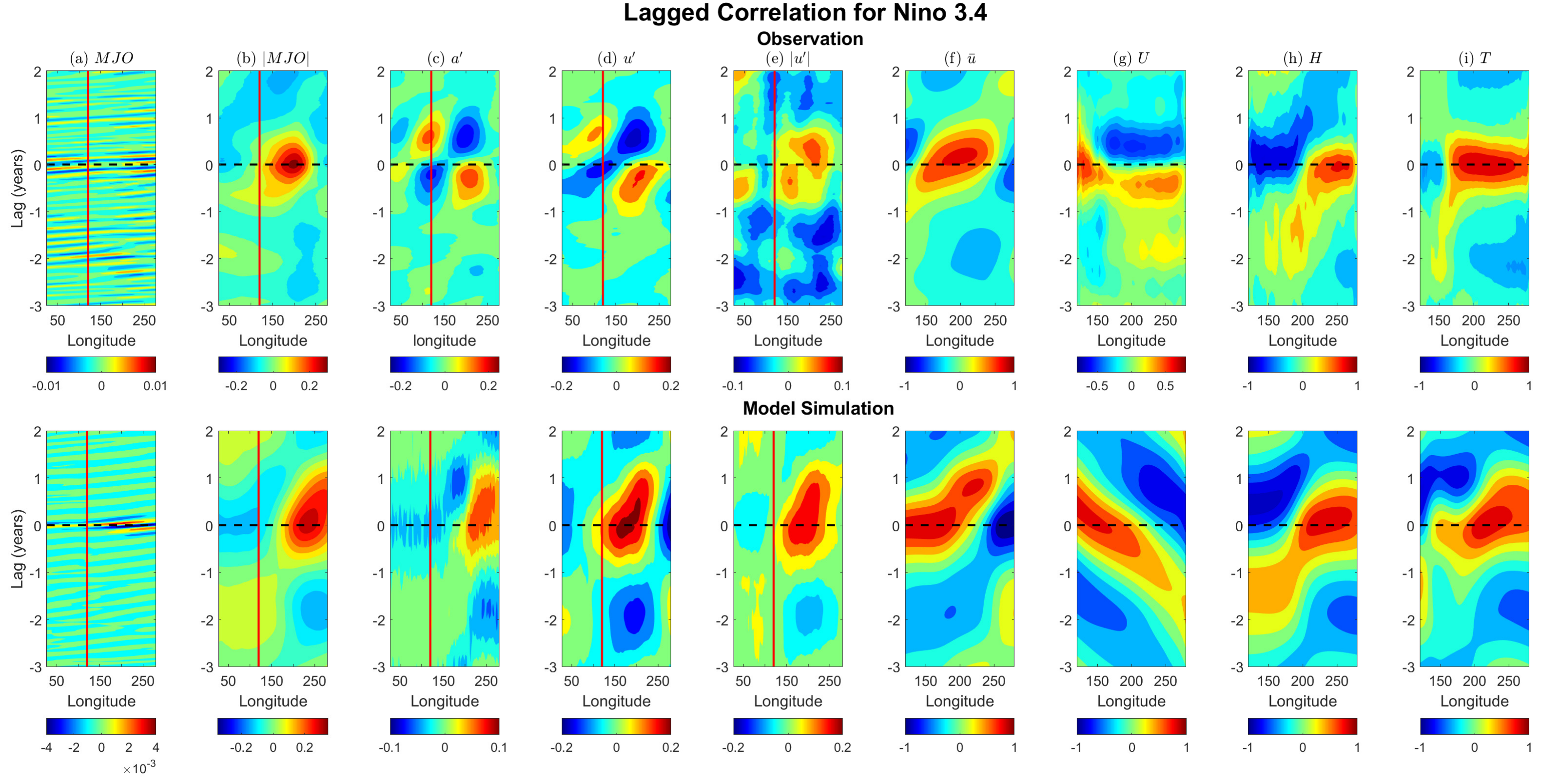}
    \caption{Lagged correlation of key atmospheric and oceanic variables against the Ni\~no 3.4 index from observations (bottom row) and the model simulation (top row). Variables shown are: (a) the MJO signal $MJO$; (b) the magnitude of the MJO signal $|MJO|$ (different from Panel (a) in Figure \ref{Figure1_hov}, this $|MJO|$ is not the 1-year running mean); (c) intraseasonal convective activity $a^{\prime}$; (d) intraseasonal wind $u^{\prime}$; (e) the magnitude of the intraseasonal wind $|u^{\prime}|$; (f) interannual wind $\bar{u}$; (h) zonal ocean current $U$; (i) thermocline depth $H$; and (j) SST. The x-axis represents longitude, and the y-axis represents lag time in years. Negative lags indicate variables precede Ni\~no 3.4 index changes, while positive lags indicate variables follow Ni\~no 3.4 index changes. Red vertical lines mark the Western Pacific edge, and black dashed horizontal lines indicate zero lag.}
    \label{Lagcorrelation_TE}
\end{figure}

Figure \ref{Lagcorrelation_TE} shows the lagged correlation between the SSTs in the Ni\~no 3.4 region and the key oceanic and atmospheric variables in the model and observations. Overall, the model results are highly consistent with observations, which demonstrates the model's ability to capture the complex relationship between MJO and ENSO. Specifically, both the model and observations indicate that the Ni\~no 3.4 index has a strong relationship with the amplitude of the MJO rather than its phase, which is consistent with the results in Figure \ref{Figure1_hov}. Furthermore, the peak positive correlations of $|MJO|$, $u'$, and $|u'|$ against SSTs propagate eastward across the Pacific as the lag approaches and crosses zero. This shift indicates that the atmospheric dynamics in the Western Pacific drive changes in SSTs, while SSTs in turn influence changes in atmospheric dynamics further east. These correlational relationships correspond to two key features of the MJO-ENSO coupled system: enhanced MJO activity and WWEs tend to occur prior to El Ni\~no onset, and ENSO-related sea surface warming often drives MJO peak activity eastward. Additionally, the correlations of zonal advection and thermocline depth against the Ni\~no 3.4 index highlight the model's skill in capturing the oceanic dynamics of ENSO. Specifically, changes in the zonal ocean current in the Central Pacific are positively correlated with and precede changes in SSTs, demonstrating the role of CP zonal advection in the formation of CP events. Finally, changes in thermocline depth in the western and central Pacific lead to SST changes, reflecting the slow adjustment of the thermocline, triggering the turnabout of ENSO.

\begin{figure}[H]
\centering
    \hspace*{-0cm}\includegraphics[width=1\textwidth]{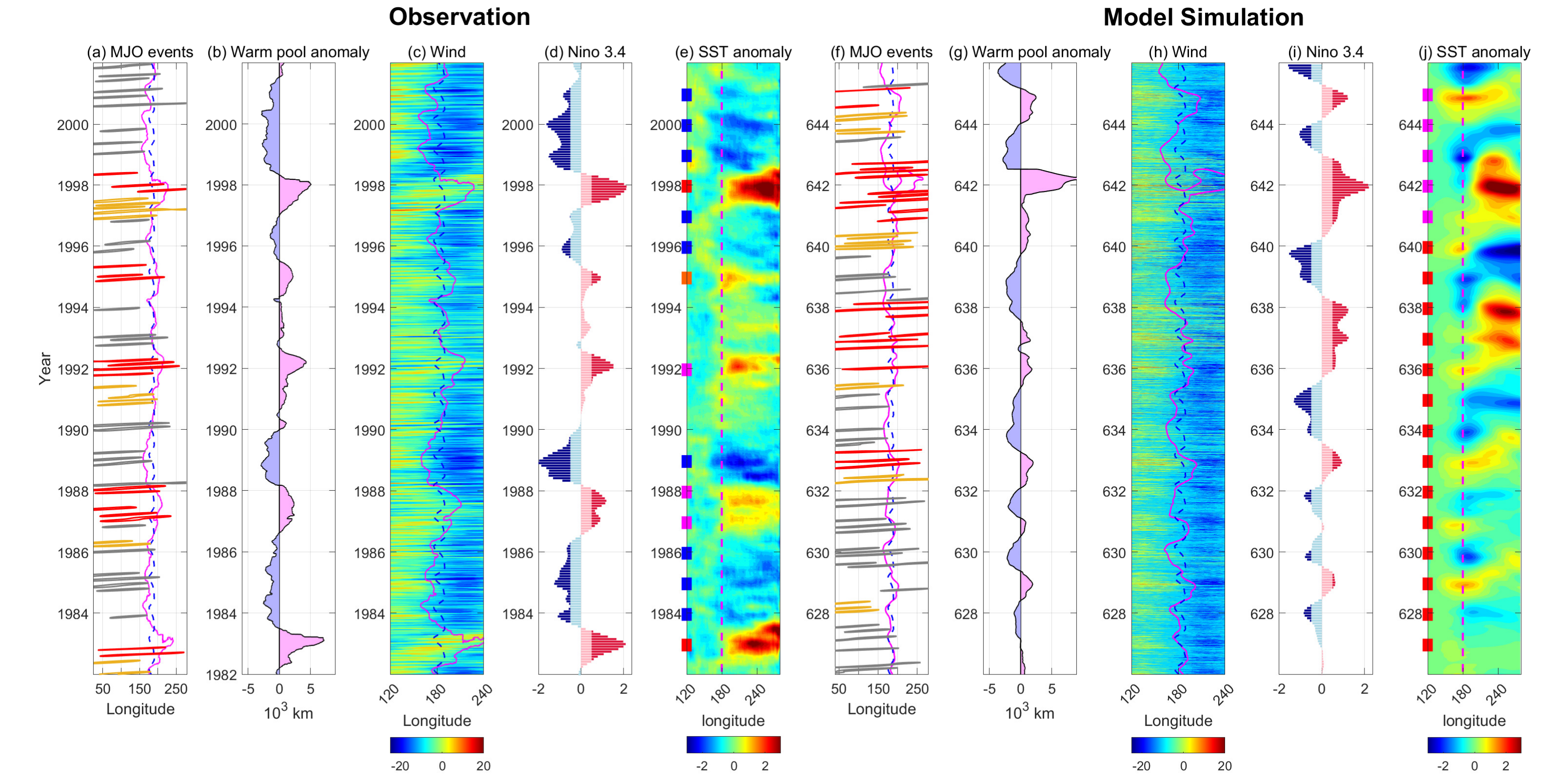}
    \caption{Comparison of observed (panels (a)--(d)) and simulated (panels (f)--(j)) MJO events, the eastern edge of the warm pool, zonal wind, and SSTs. Panels (a) and (f): the eastern edge of equatorial warm pool (28.5$^{\circ}$C; magenta) averaged from 5$^{\circ}$S to 5$^{\circ}$N and its climatological position (blue dashed) for 20 years of data. The MJO events occurring within 9 months prior to an El Ni\~no onset are shown in orange, during El Ni\~no in red, and the remaining events in gray. Panels (b) and (g): the Warm Pool Eastward Extension (WPEE; defined as the eastward displacement of the 28.5$^{\circ}$C isotherm from its climatological position), where its negative and positive anomalies ($10^3$ km) are shaded in light purple and light pink, respectively. Panels (c) and (h): total equatorial surface zonal winds with the background component. Panels (d) and (i): Ni\~no-3.4 SST anomaly (3-month average), where the anomalies $>$ 0.5$^{\circ}$C are shown in red and those $<$-0.5$^{\circ}$C are shown in blue. Panels (e) and (j): SST anomaly.}
    \label{Warmpool_test}
\end{figure}

Figure \ref{Warmpool_test} shows the spatiotemporal patterns of MJO events, the eastern edge of the warm pool, zonal wind, and SSTs. Here, the data is analyzed in a manner similar to the observational study \cite{jauregui2024mjo} to further assess the model's performance in representing the interactions between MJO and ENSO. The model successfully reproduces the intensification of MJO activity leading up to and throughout El Ni\~no events, as shown in orange and red in panels (a) and (f). For extreme events, more MJO activity is observed in the lead-up (observation year 1997 and model year 642). Additionally, both observations and the model evidence the theory that MJO may contribute to El Ni\~no development, by inducing westerly wind bursts (panels (c) and (h)), which trigger down-welling Kelvin waves that extend the warm pool eastward (panels (b) and (g)). The model also generates warming in the Ni\~no 3.4 region following the Warm Pool Eastward Extension (WPEE). Thus, the atmospheric variability driven by latent heat and the wind stress forcing on the zonal ocean current are highly effective in representing the interactions between MJO and ENSO.

\begin{figure}[H]
\centering
    \hspace*{-0cm}\includegraphics[width=1\textwidth]{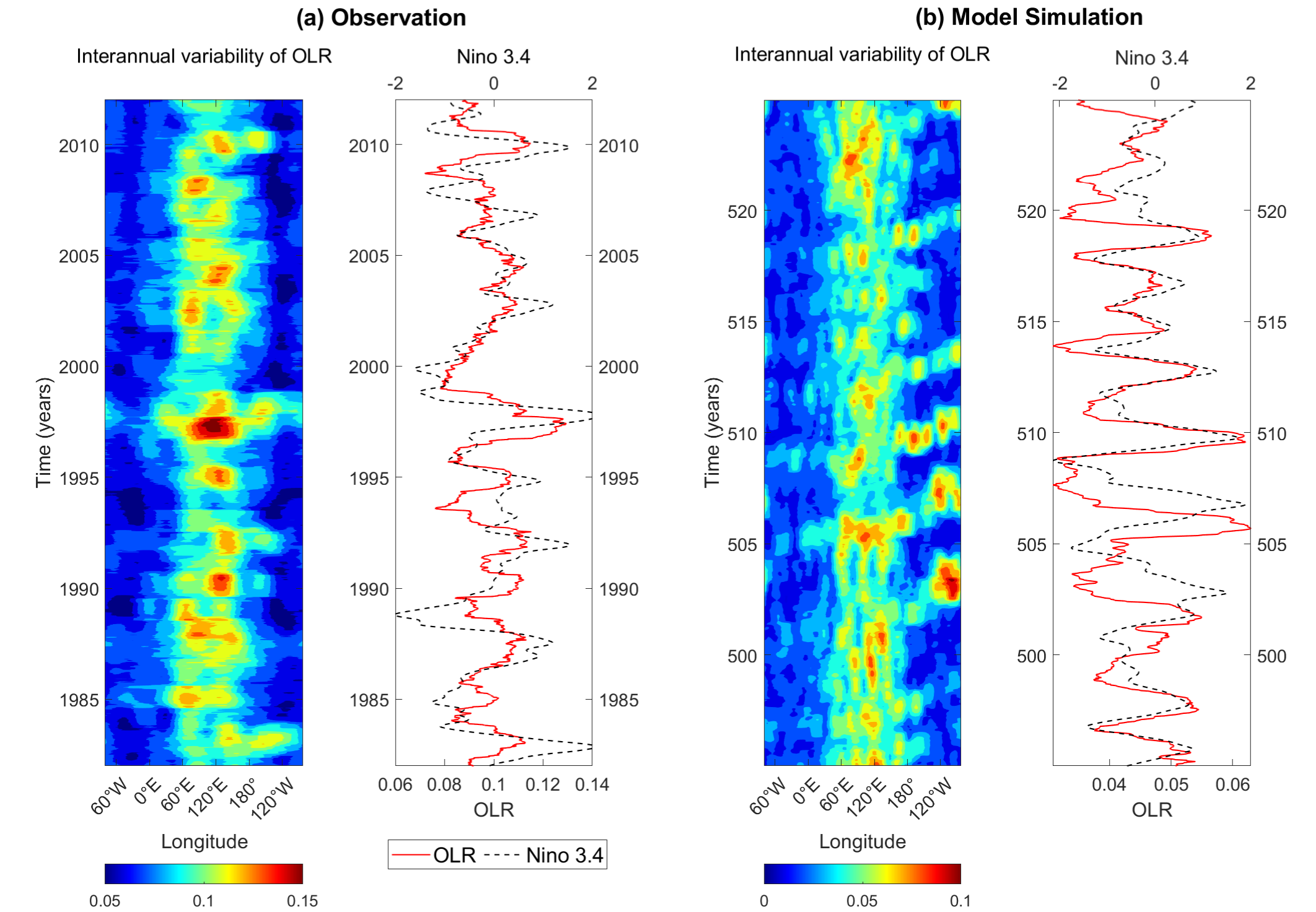}
    \caption{Comparison of the interannual amplitude of intraseasonal OLR variability in observations (panel (a)) and the model simulation (panel (b)). The left plot in each panel shows the interannual OLR amplitude in the entire global tropical strip, centered on the major region of variance (the abscissa extends around the globe, broken at the South American coast at 80$^{\circ}$W). The right plots in each panel show time series of the OLR interannual amplitude averaged over the Western Pacific (150$^{\circ}$E–180$^{\circ}$; red solid line) in comparison with the Ni\~no 3.4 index (black dashed line).}
    \label{OLR_hov}
\end{figure}

Figure \ref{OLR_hov} shows the Hovmoller diagram and time series for the interannual amplitude of outgoing long-wave radiation (OLR) alongside the temporal evolution of the Ni\~no 3.4 index. Here, the modeled intraseasonal convective activity $a'$ is used as a surrogate for OLR, as they have a very strong negative correlation \cite{majda2009skeleton, stechmann2015identifying}. The data are processed in a manner similar to the observational study \cite{kessler2001eof} to validate the model's performance in generating MJO convective patterns and how they relate to ENSO. That is, the observational and model OLR data are filtered to the intraseasonal band (30-90 day periods), and a spatial filter retaining wave numbers $|k| \in [0,10]$ is applied to remove small-scale noise. Finally, the interannual amplitude of OLR is obtained by applying a 1-year moving average to the square of the filtered data and then taking the square root.

The model can generate realistic spatial dynamics of the interannual OLR amplitude (left plots in panels (a) and (b)). Particularly, OLR variability in both observations and model simulations exhibits consistent maxima near the eastern edge of the Indian Ocean at $90^\circ$E and intermittent maxima that coincide with El Ni\~no events, extending into the Pacific. Furthermore, the model reproduces the close tie between OLR in the Western Pacific and the Ni\~no 3.4 index (right plots in panels (a) and (b)), indicating enhanced convective activity in the Western Pacific during El Ni\~no events.

\subsection{Sensitivity tests}

\subsubsection{Impact of decadal variability}

Decadal variability plays a critical role in producing realistic ENSO complexity and diversity by modulating the coupling between the ocean and the atmosphere. Depending on the value of $I$, different oceanic patterns will emerge. The stochastic dynamics governing the decadal variable allow the model to produce a realistic irregular alternation between periods dominated by EP and CP events. To isolate the effects of different Walker circulation regimes, we conduct sensitivity experiments with constrained $I$ values: (1) $I \in [0.25, 0.75]$ representing moderate variability, and (2) $I = 0.5$ representing a fixed intermediate state.

When the decadal variability is constrained to $I \in [0.25, 0.75]$, the model produces fewer EP events compared to the full range. The occurrence frequencies become: 28.5 ± 7.0 El Ni\~no events per 71 years, with EP events decreasing from $18.6 \pm 6.8$ to $12.9 \pm 7.3$ and CP events maintaining $15.6 \pm 2.8$. Most notably, extreme El Ni\~no events are reduced to only $1.0 \pm 1.3$ per 71 years, compared to $3.4 \pm 2.7$ in the standard model formulation and $4$ in observations. This reduction occurs because the constrained range limits the model's ability to reach the weak Walker circulation states that favor the development of extreme EP events through enhanced thermocline feedback. When $I$ is fixed at 0.5, the model generates $25.1 \pm 5.7$ El Ni\~no events per 71 years, with even fewer EP events ($10.1 \pm 5.2$) while CP events are maintained ($15.0 \pm 1.8$). Furthermore, extreme El Ni\~no events become even more rare ($0.3 \pm 0.6$). These findings are consistent with observational evidence that the most extreme El Ni\~no events, such as those in 1982-1983, 1997-1998, and 2015-2016, occurred during periods characterized by relatively weak Walker circulation states. These model results demonstrate that decadal variability is essential for realistic event frequencies.

\subsubsection{Role of state-dependent noises in the atmospheric dynamics}

Recall from \eqref{Eq:Atm_intra_1} that the noise coefficients $\hat{\sigma}_q$ and $\hat{\sigma}_a$ are state-dependent. As discussed in Section \ref{Subsec:Interannual_Intraseasonal}, these state-dependent noises play a crucial role in allowing interannual variabilities to modulate the dynamics and statistics of the intraseasonal components. To examine the necessity of adopting state-dependent noises, we perform additional experiments replacing them with additive noises. In such a way, the feedback from interannual variables to the intraseasonal components via stochastic forcing is eliminated. When performing these experiments, we empirically tune the strengths of these additive noises to maintain the amplitudes and patterns of the MJO and ENSO events as much as possible.

When the state-dependent noise in the moisture process is replaced with additive noise, fundamental changes appear in the model dynamics and statistics. The wind variability becomes increasingly random. This is because the wind no longer varies in response to SST variations. Therefore, it lacks any significant structure on the interannual timescale. Consequently, extreme EP El Ni\~no events are rarer, and the Ni\~no 3 SST PDF becomes quite Gaussian.

Next, we modify the noise coefficient in the convective activity equation from state-dependent to additive. With such a change, the number of extreme MJO events will decrease, which has also been found in \cite{stachnik2015evaluating, ogrosky2015mjo}. While the SST variability and the amplitude of extreme El Ni\~nos are somewhat preserved under appropriately tuned additive noise, the overall SST pattern becomes more regular than the case with state-dependent noise. In addition, the amplitudes of the EP La Ni\~na events are generally underestimated, causing the EP SST mean value to increase. This is likely because the additive noise lacks the mechanisms that suppress wind and convective activity after El Ni\~no events, as the feedback from SST to convective activity is eliminated.

\section{Conclusion and Discussions}\label{Sec:Conclusion}
This paper presents a simple stochastic intermediate coupled MJO-ENSO model designed to understand their bidirectional feedback and its role in modulating ENSO complexity. The model integrates multiscale processes, bridging intraseasonal (MJO), interannual (ENSO), and decadal (Walker circulation) time scales. Numerical simulations show that the model can reproduce several crucial observed MJO and ENSO features, including non-Gaussian statistics, seasonal cycles, energy spectra, and spatial event patterns. Furthermore, the model captures essential MJO-ENSO interactions, such as warm pool edge extension, convective activity adjustments that modulate SSTs, and ENSO's dependence on MJO-driven easterly and westerly wind anomalies.

Beyond understanding the spatiotemporal evolution of the MJO-ENSO dynamics, the model provides several broader applications.
First, the model facilitates the predictability studies of MJO and ENSO, including studying the statistical predictability \cite{fang2023quantifying}. Notably, the model allows for the examination of how the MJO affects ENSO forecast skill. Therefore, the model can help determine whether explicitly resolving the MJO in atmospheric models is essential or if a simplified wind parameterization suffices for the forecast purpose.
Second, the model applies to long-term climate analysis. Due to its computational efficiency compared to operational models, it can generate extended time series. These long time series advance the study of rare and extreme events that have occurred more frequently in recent decades. They also enable the study of the statistical response of the MJO and ENSO to climate perturbations. In addition, these long simulations facilitate various machine learning applications.
Third, the intermediate coupled model developed here can synergize with operational models. Despite the lower resolution, the statistical accuracy of the model makes it valuable for correcting certain errors in operational models through multi-model data assimilation \cite{wang2025dual}. When the simulation from this intermediate coupled model is integrated with the operational models, the combined system can produce high-resolution, statistically robust simulations. This has the potential to bridge the gap between conceptual understanding and operational forecasting.

\section{Open Research}


The dataset supporting this research is available on Zenodo \cite{zhang2025data}. The MATLAB code used for data processing and figure generation is preserved on Zenodo \cite{zhang2025code} and is actively developed on GitHub (\url{https://github.com/ylzhang2447/MJO-ENSO}).

\section{Acknowledgments}
The research of N.C. is funded by the Office of Naval Research N00014-24-1-2244 and the Army Research Office W911NF-23-1-0118. C.M. is  supported as a research assistant under the first grant.

\section{Appendix}

\subsection{Meridional truncation for the full system}\label{Appendix:Meri}

The atmosphere is defined over the entire equatorial belt $0\leq x\leq L_A$, while the ocean is defined on only a portion $0 \leq x \leq L_O$, representing the Pacific Ocean, with $L_O < L_A$. The atmosphere has periodic boundary conditions, $\bar{u}(0,y,t) = \bar{u}(L_A,y,t)$ (the same for other variables). The ocean has reflection boundary conditions, $\int_{-\infty}^{+\infty}U(0,Y,t)\mathrm{d} Y = 0$ and $U(L_O,Y,t) =0$.

The full coupled MJO-ENSO model \eqref{Eq:Atm_intra_1}--\eqref{Eq:Couplings_4} is solved by projecting it meridionally using parabolic cylinder functions \cite{majda2003introduction}. This is similar to Fourier analysis, where the parabolic cylinder functions are analogs of the Fourier basis, while the projection coefficients (now are functions of $x$ and $t$) mimic the Fourier coefficients. The reason parabolic cylinder functions appear in the coupled model is due to the presence of the coordinate variables $y$ and $Y$ in the system, which prevents the use of Fourier bases in the meridional projections. Since the resulting system corresponding to each meridionally projected mode depends only on $x$ and $t$, it is computationally much easier to handle. The parabolic cylinder functions for the atmosphere and ocean differ in their meridional extent, or Rossby radius, but share the same structure. The first and second parabolic functions for atmosphere are $\phi_0(y) = (\pi)^{-1/4} \exp(-y^2/2)$ and $\phi_1(y) = (4\pi)^{-1/4}(2y^2-1)\exp(-y^2/2)$. The oceanic parabolic functions are similar to the atmospheric ones but depend on $Y$ instead, with $Y = y/\sqrt{c}$, and are denoted $\psi_m(Y)$. As the atmosphere and ocean communicate, the projected variables in the atmosphere will map to the ocean and vice versa. Due to the different meridional extents, mapping coefficients from the atmosphere to the ocean and from the ocean to the atmosphere need to be introduced. Using the leading meridional modes as an example, these mapping coefficients are given by
\begin{equation}
\begin{aligned}
    \chi_A & = \int_{-\infty}^{+\infty}\phi_0(y)\phi_0(y/\sqrt{c})\mathrm{d} y,\\
    \chi_O & = \int_{-\infty}^{+\infty}\psi_0(Y)\psi_0(\sqrt{c}Y)\mathrm{d} y.
\end{aligned}
\end{equation}

In this paper, the model is projected only to the leading meridional basis. The atmosphere variables are written as $\left\{u, v, \theta, E_q\right\}(x, y, t)=\left\{u, v, \theta, \chi_A E_q\right\}(x, t) \phi_0(y)$ and the ocean variables are given by $\left\{U, V, H, \tau_x\right\}(x, Y, t)=\left\{U, V, H, \chi_O \tau_x\right\}(x, t) \psi_0(Y)$. When transforming the physical variables to the characteristic variables for the projected system, atmospheric Kelvin and Rossby waves, $K_A$ and $R_A$, as well as oceanic Kelvin and Rossby waves, $K_O$ and $R_O$, are triggered. In addition, there is $\left\{T, E_q\right\}(x, Y, t)=\left\{T, E_q\right\}(x, t) \psi_0(Y)$ for SST. Note that the same notation is used for variables after the projection, namely, on the right-hand side of the above relationships, but they are functions depending only on $(x,t)$.

The coupled model in \eqref{Eq:Atm_intra_1}--\eqref{Eq:Couplings_4}, after being projected to the first parabolic cylinder functions reads:

Intraseasonal atmosphere:

\begin{equation}\label{Eq:m1}
\begin{aligned}
\left(\partial_t+d_u\right) K_A^{\prime}+\partial_x K_A^{\prime} & =-\bar{H} a^{\prime} / 2, \\
\left(\partial_t+d_u\right) R_A^{\prime}-\partial_x R_A^{\prime} / 3 & =-\bar{H} a^{\prime} / 3, \\
\left(\partial_t+d_q\right) Q^{\prime}+\bar{Q}\left(\partial_x K_A^{\prime}-\partial_x R_A^{\prime} / 3\right) & =-\bar{H} a^{\prime}(1-\bar{Q} / 6) + \sigma_q(E_q,s_q,I)\dot{W}_q,\\
\left(\partial_t+d_a\right) A^{\prime} & =\Gamma Q^{\prime}(\bar{A}+A^{\prime}) + \sigma_a(\bar{A},A',Q')\dot{W}_a.
\end{aligned}
\end{equation}

Interannual atmosphere:
\begin{equation}\label{Eq:m2}
\begin{aligned}
\partial_x \bar{K}_A & =-\chi_A\left(E_q-<E_q>\right)(2-2 \bar{Q})^{-1}, \\
-\partial_x \bar{R}_A / 3 & =-\chi_A\left(E_q-<E_q>\right)(3-3 \bar{Q})^{-1},\\
\bar{K}_A(0, \tau) & =\bar{K}_A\left(L_A, \tau\right), \\
\bar{R}_A(0, \tau) & =\bar{R}_A\left(L_A, \tau\right), \\
\bar{H} \bar{a} & =\left(E_q - <E_q>+s^q-\bar{Q} s^\theta\right) /\left[\delta_A(1-\bar{Q})\right].
\end{aligned}
\end{equation}

Ocean:
\begin{equation}\label{Eq:m3}
\begin{aligned}
\partial_t K_O+c_1\partial_x K_O & = \chi_O c_1 \tau_x / 2, \\
\partial_t R_O-\left(c_1/ 3\right) \partial_x R_O & =-\chi_O c_1 \tau_x / 3, \\
K_O(0, t) & =r_W R_O(0, t), \\
R_O\left(L_O, t\right) & =r_E K_O\left(L_O, t\right).
\end{aligned}
\end{equation}

SST:
\begin{equation}\label{Eq:m4}
\partial_t T=-c_1\zeta E_q + c_1  \eta_1\left(K_O+R_O\right)+ \kappa c_1   \eta_2\left(K_O-R_O\right).
\end{equation}

Couplings:
\begin{equation}\label{Eq:m5}
\begin{aligned}
& \tau_x=\gamma\left(\bar{K}_A-\bar{R}_A+K_A^{\prime} - R_A^{\prime}\right), \\
& E_q=\alpha_q T.
\end{aligned}
\end{equation}

Once these waves are solved, the physical variables can be reconstructed,
\begin{equation}\label{Eq:Reconstruction}
\begin{aligned}
& u=\left(K_A-R_A\right) \phi_0+\left(R_A / \sqrt{2}\right) \phi_2, \\
& \theta=-\left(K_A+R_A\right) \phi_0-\left(R_A / \sqrt{2}\right) \phi_2, \\
& U=\left(K_O-R_O\right) \psi_0+\left(R_O / \sqrt{2}\right) \psi_2, \\
& H=\left(K_O+R_O\right) \psi_0+\left(R_O / \sqrt{2}\right) \psi_2,
\end{aligned}
\end{equation}
where $\phi_2$ and $\psi_2$ are the third meridional bases of atmosphere and ocean, respectively, resulting from the meridional expansion of the parabolic cylinder functions. Note that the mean latent heat $<E_q>$  is subtracted in order for the latent heat forcing to maintain a zonal equatorial mean of zero in the absence of atmospheric dissipation \cite{majda2003systematic,stechmann2014walker}.

All reference scales and parameters are listed in tables \ref{tab:table1} and \ref{tab:table2}.

\begin{table}[H]
\centering
\begin{tabular}{|l|l|l|}
\hline Variable & Unit & Unit value \\
\hline $x$ zonal axis & $[y] / \delta$ & $15,000 \mathrm{~km}$ \\
\hline y atmospheric meridional axis & $\sqrt{c_A / \beta}$ & $1,500 \mathrm{~km}$ \\
\hline $Y$ oceanic meridional axis & $\sqrt{c_o / \beta}$ & 330 km \\
\hline $t$ interannual time axis & [$t$] & 34 days \\
\hline $\bar{u}$ interannual zonal wind speed & $\delta c_A$ & $5 \mathrm{~m}\cdot\mathrm{s}^{-1}$ \\
\hline $u'$ intraseasonal zonal wind speed & $\delta c_A$ & $5 \mathrm{~m}\cdot\mathrm{s}^{-1}$ \\
\hline $v$ meridional wind speed & $\delta[u]$ & $0.5 \mathrm{~m}\cdot\mathrm{s}^{-1}$ \\
\hline $\bar{\theta}$ interannual potential temperature & $15 \delta$ & 1.5 K \\
\hline $\theta'$ intraseasonal potential temperature & $15 \delta$ & 1.5 K \\
\hline $E_q$ latent heating & $[\theta] /[t]$ & $0.45 \mathrm{~K}\cdot{\text {day }}{ }^{-1}$ \\
\hline U zonal current speed & $c_O \delta_O$ & $0.25 \mathrm{~m}\cdot\mathrm{s}^{-1}$ \\
\hline $V$ meridional current speed & $\delta \sqrt{c}[U]$ & $0.56 \mathrm{~cm}\cdot \mathrm{s}^{-1}$ \\
\hline $H$ thermocline depth & $H_O \delta_O$ & 20.8 m \\
\hline $T$ sea surface temperature & [$\theta$] & 1.5 K \\
\hline
\end{tabular}
\caption{Model variables, definitions and units.}
\label{tab:table1}
\end{table}

\begin{table}[H]
    \centering
    \begin{tabular}{|p{8cm}|p{8cm}|}
    \hline Parameter & Value \\
    \hline $\epsilon$ froude number & 0.278\\
    \hline $\delta$ long-wave scaling & 0.1\\
    \hline $\delta_O$ arbitrary constant & 0.1\\
    \hline $\delta_A$ interannual atmospheric coupling factor & 0.1\\
    \hline$c_A$ atmospheric phase speed & $50 {~m\cdot s}^{-1}$ \\
    \hline$c_O$ oceanic phase speed & $\sqrt{g^{\prime} H_O} = 2.5 {~m\cdot s}^{-1}$ \\
    \hline $c$ ratio of ocean/atmosphere phase speed & $c_O/c_A = 0.05$ \\
    \hline $c_1$ modified ratio of phase speed & $c/\epsilon = 0.18$ \\
    \hline $\beta$ beta-plane parameter & $2.28~10^{-11}~m^{-1}~s^{-1}$\\
    \hline $g'$ reduced gravity & $0.03~m\cdot s^{-2}$\\
    \hline $H_O$ mean thermocline depth & 208 $m$\\
    \hline $\rho_O$ ocean density & 1000 $kg\cdot m^{-3}$\\
    \hline $\bar{H}$ convective heating rate factor & 22 \\
    \hline $\bar{Q}$ mean vertical moisture gradient & 0.9 \\
    \hline$\chi_A$ atmospheric meridional projection coefficient & 0.31 \\
    \hline$\chi_O$ oceanic meridional projection coefficient & 1.38 \\
    \hline $L_A$ equatorial belt length & $8 / 3$ \\
    \hline $L_O$ equatorial Pacific length & 1.2 \\
    \hline $\alpha_q$ latent heating factor & $\alpha_q=q_c q_e \exp \left(q_e \bar{T}\right) / \tau_q$ \\
    \hline $\bar{T}$ mean SST & $16.6\left(\right.$which is $\left.25^{\circ} \mathrm{C}\right)$ \\
    \hline $q_c$ latent heating multiplier coefficient & 8.75 \\
    \hline $q_e$ latent heating exponential coefficient & 0.093 \\
    \hline $\tau_q$ latent heating adjustment rate & 15 \\
    \hline $\gamma$ wind stress coefficient & 5.876 \\
    \hline$\kappa(I)$ zonal advection feedback strength coefficient & $0.1+ 1.2I$ \\
    \hline $r_W$ western boundary reflection coefficient& 0.5 \\
    \hline $r_E$ eastern boundary reflection coefficient & 1  \\
    \hline $\zeta$ latent heating exchange coefficient & $20.385\times\beta_1\times \beta_2$ \\
    \hline $\beta_1(T)$ state dependent component in $\zeta$ & $\beta_1(T)={1.8}-\eta_2 / {3}+\left(0.2+\left|T_C+0.4\right| \times \eta_2\right)^2 / {5}$ \\
    \hline $\beta_2(t)$ seasonal dependent component in $\zeta$ & $\beta_2(t)$ = 1 + {0.4} $\sin (2 \pi t)$  + {0.0877} $\sin (2 \pi(t+1 / 12)) \eta_1 $ + {0.1389} $\sin (2 \pi t{-2/12}) \eta_2$\\
    \hline $c_2$ mean correction coefficient & 0.1 \\
    \hline $\eta_1$ profile of thermocline feedback & $\eta_1(x)=1.235+\left(1.045 \times \tanh \left(7.5\left(x-L_O / 3\right)\right)\right)$\\
    \hline $\eta_2$ profile of zonal advective feedback & $\eta_2(x)=$ $\max$ (0,4 $\exp (-(x-L_O /(7 / 3))^2 / 0.1)\times$ 0.9) \\
    \hline $\lambda$ damping of decadal variability & {0.0189} (which is {5} years$^{-1}$) \\
    \hline $m$ mean of $I$  & 0.5\\
    \hline$\Gamma$ convective growth/decay rate & 16.6 \\
    \hline$d_a, d_q, d_\theta, \lambda_a$ atmosphere dissipations & 1.1 (which is 1 month$^{-1}$) \\
    \hline $s^q$ external moistening source & 0.88 + $2.64/({1 + \exp(-10x_p + 6)})$ - $2.64/(1 + \exp(-10x_p - 3))$,\quad $x_p = \begin{cases}
    x, & \text{if } x \in [0, L_A/2] \\
    x - L_A, & \text{if } x \in (L_A/2, L_A]
    \end{cases}$ \\
    \hline$s^\theta$ external cooling source & $s^\theta=s^q$\\
    \hline $\sigma_q$ the noise of $Q'$ & $\sigma_q (\xi) = 4\times(0.77 + \tanh(0.23\xi+1))$\\
    \hline $\xi$ composite parameter in the noise of $Q'$ & $\xi = \max ( (\beta_3 E_q - <\beta_3E_q>) + (1- I)s^q,0.1 )$, $\beta_3 = 23+7.368\eta_1$\\
    \hline $\sigma_a$ the noise of $A'$ & $\sigma_a = \sqrt{166(\bar{A}+A')|Q'|}$\\
    \hline
    \end{tabular}
    \caption{Model parameter values for the meridionally truncated model \eqref{Eq:m1}-\eqref{Eq:m5}.}
    \label{tab:table2}
\end{table}

\subsection{Details of the linear eigenmode analysis of the interannual component}\label{Sec:LinearSolution}

The linear solutions in Figure \ref{fig:Figure_LinearSolution} are computed from the deterministic coupled system \eqref{Eq:Atm_inter}--\eqref{Eq:SST2} with all stochastic triggering effects from the intraseasonal wind removed ($\tau_x = \gamma \bar{u}$). Decadal variability $I$, and thus the Walker circulation modulation parameter $\kappa(I)$, is fixed to be a constant. Nonlinear components in the damping term $\zeta$ are omitted. 
Panel (a) shows the EP-dominant solution with $I =1/6$ ($\kappa(I) = 0.3$), reducing the strength of zonal advection feedback to highlight thermocline-driven dynamics. In contrast, Panel (b) shows the CP-dominant solution with $I = 2/3$ ($\kappa(I) = 0.9$), strengthening zonal advection feedback. Note that some parameters ($\alpha_q = 0.2043, \gamma = 6.529,$ and $\zeta = 8.7$) are adjusted from the full model to compensate for the lack of intraseasonal wind forcing.

\bibliographystyle{unsrt}
\bibliography{references}

\begin{thebibliography}{100}

\bibitem{mcphaden2006enso}
Michael~J McPhaden, Stephen~E Zebiak, and Michael~H Glantz.
\newblock {ENSO} as an integrating concept in earth science.
\newblock {\em science}, 314(5806):1740--1745, 2006.

\bibitem{ropelewski1987global}
Chester~F Ropelewski and Michael~S Halpert.
\newblock {Global and regional scale precipitation patterns associated with the El Ni{\~n}o Southern Oscillation}.
\newblock {\em Monthly weather review}, 115(8):1606--1626, 1987.

\bibitem{dai2000global}
Aiguo Dai and TML Wigley.
\newblock {Global patterns of ENSO-induced precipitation}.
\newblock {\em Geophysical Research Letters}, 27(9):1283--1286, 2000.

\bibitem{ashok2007nino}
Karumuri Ashok, Swadhin~K Behera, Suryachandra~A Rao, Hengyi Weng, and Toshio Yamagata.
\newblock {El Ni{\~n}o Modoki and its possible teleconnection}.
\newblock {\em Journal of Geophysical Research: Oceans}, 112(C11), 2007.

\bibitem{capotondi2015understanding}
Antonietta Capotondi, Andrew~T Wittenberg, Matthew Newman, Emanuele Di~Lorenzo, Jin-Yi Yu, Pascale Braconnot, Julia Cole, Boris Dewitte, Benjamin Giese, Eric Guilyardi, et~al.
\newblock {Understanding {ENSO} diversity}.
\newblock {\em Bulletin of the American Meteorological Society}, 96(6):921--938, 2015.

\bibitem{timmermann2018nino}
Axel Timmermann, Soon-Il An, Jong-Seong Kug, Fei-Fei Jin, Wenju Cai, Antonietta Capotondi, Kim~M Cobb, Matthieu Lengaigne, Michael~J McPhaden, Malte~F Stuecker, et~al.
\newblock {El Ni{\~n}o--Southern Oscillation complexity}.
\newblock {\em Nature}, 559(7715):535--545, 2018.

\bibitem{kao2009contrasting}
Hsun-Ying Kao and Jin-Yi Yu.
\newblock Contrasting eastern-{P}acific and central-{P}acific types of {ENSO}.
\newblock {\em Journal of Climate}, 22(3):615--632, 2009.

\bibitem{kim2012statistical}
Jin-Soo Kim, Kwang-Yul Kim, and Sang-Wook Yeh.
\newblock Statistical evidence for the natural variation of the central {Pacific} {El Ni{\~n}o}.
\newblock {\em Journal of Geophysical Research: Oceans}, 117(C6), 2012.

\bibitem{chen2008nino}
Dake Chen and Mark~A Cane.
\newblock El {N}i{\~n}o prediction and predictability.
\newblock {\em Journal of Computational Physics}, 227(7):3625--3640, 2008.

\bibitem{jin2008current}
Emilia~K Jin, James~L Kinter, Bin Wang, C-K Park, I-S Kang, BP~Kirtman, J-S Kug, A~Kumar, J-J Luo, J~Schemm, et~al.
\newblock Current status of {ENSO} prediction skill in coupled ocean--atmosphere models.
\newblock {\em Climate Dynamics}, 31(6):647--664, 2008.

\bibitem{barnston2012skill}
Anthony~G Barnston, Michael~K Tippett, Michelle~L L'Heureux, Shuhua Li, and David~G DeWitt.
\newblock Skill of real-time seasonal {ENSO} model predictions during 2002--11: {I}s our capability increasing?
\newblock {\em Bulletin of the American Meteorological Society}, 93(5):631--651, 2012.

\bibitem{hu2012analysis}
Zeng-Zhen Hu, Arun Kumar, Bhaskar Jha, Wanqiu Wang, Bohua Huang, and Boyin Huang.
\newblock An analysis of warm pool and cold tongue {E}l {N}i{\~n}os: {A}ir--sea coupling processes, global influences, and recent trends.
\newblock {\em Climate Dynamics}, 38(9):2017--2035, 2012.

\bibitem{zheng2014asymmetry}
Fei Zheng, Xianghui Fang, Jin-Yi Yu, and Jiang Zhu.
\newblock Asymmetry of the {B}jerknes positive feedback between the two types of {E}l {N}i{\~n}o.
\newblock {\em Geophysical Research Letters}, 41(21):7651--7657, 2014.

\bibitem{fang2015cloud}
Xianghui Fang, Fei Zheng, and Jiang Zhu.
\newblock The cloud-radiative effect when simulating strength asymmetry in two types of {E}l {N}i{\~n}o events using {CMIP5} models.
\newblock {\em Journal of Geophysical Research: Oceans}, 120(6):4357--4369, 2015.

\bibitem{sohn2016strength}
Soo-Jin Sohn, Chi-Yung Tam, and Hye-In Jeong.
\newblock How do the strength and type of {ENSO} affect {SST} predictability in coupled models.
\newblock {\em Scientific Reports}, 6(1):1--8, 2016.

\bibitem{santoso2019dynamics}
Agus Santoso, Harry Hendon, Andrew Watkins, Scott Power, Dietmar Dommenget, Matthew~H England, Leela Frankcombe, Neil~J Holbrook, Ryan Holmes, Pandora Hope, et~al.
\newblock Dynamics and predictability of {E}l {N}i{\~n}o--{S}outhern {O}scillation: an {A}ustralian perspective on progress and challenges.
\newblock {\em Bulletin of the American Meteorological Society}, 100(3):403--420, 2019.

\bibitem{bjerknes1969atmospheric}
Jakob Bjerknes.
\newblock Atmospheric teleconnections from the equatorial {Pacific}.
\newblock {\em Monthly Weather Review}, 97(3):163--172, 1969.

\bibitem{madden1971detection}
Roland~A Madden and Paul~R Julian.
\newblock {Detection of a 40--50 day oscillation in the zonal wind in the tropical Pacific}.
\newblock {\em Journal of Atmospheric Sciences}, 28(5):702--708, 1971.

\bibitem{madden1994observations}
Roland~A Madden and Paul~R Julian.
\newblock {Observations of the 40--50-day tropical oscillation—A review}.
\newblock {\em Monthly weather review}, 122(5):814--837, 1994.

\bibitem{zhang2005madden}
Chidong Zhang.
\newblock {M}adden-{J}ulian {Oscillation}.
\newblock {\em Reviews of Geophysics}, 43(2), 2005.

\bibitem{woolnough2019madden}
Steven~J Woolnough.
\newblock The {M}adden-{J}ulian {O}scillation.
\newblock In {\em Sub-seasonal to seasonal prediction}, pages 93--117. Elsevier, 2019.

\bibitem{zhang2020four}
C~Zhang, {\'A}F~Adames, B~Khouider, B~Wang, and D~Yang.
\newblock Four theories of the {M}adden-{J}ulian {O}scillation.
\newblock {\em Reviews of Geophysics}, 58(3):e2019RG000685, 2020.

\bibitem{puy2016modulation}
Martin Puy, J{\'e}r{\^o}me Vialard, Matthieu Lengaigne, and Eric Guilyardi.
\newblock {Modulation of equatorial Pacific westerly/easterly wind events by the {Madden--Julian Oscillation} and convectively-coupled Rossby waves}.
\newblock {\em Climate dynamics}, 46:2155--2178, 2016.

\bibitem{puy2019influence}
Martin Puy, J{\'e}r{\^o}me Vialard, Matthieu Lengaigne, Eric Guilyardi, Pedro~N DiNezio, Aurore Voldoire, Magdalena Balmaseda, Gurvan Madec, Christophe Menkes, and Michael~J Mcphaden.
\newblock Influence of westerly wind events stochasticity on {E}l {N}i{\~n}o amplitude: {T}he case of 2014 vs. 2015.
\newblock {\em Climate Dynamics}, 52:7435--7454, 2019.

\bibitem{tang2008mjo}
Youmin Tang and Bin Yu.
\newblock {MJO} and its relationship to {ENSO}.
\newblock {\em Journal of Geophysical Research: Atmospheres}, 113(D14), 2008.

\bibitem{mcphaden2006large}
Michael~J McPhaden, Xuebin Zhang, Harry~H Hendon, and Matthew~C Wheeler.
\newblock Large scale dynamics and {MJO} forcing of {ENSO} variability.
\newblock {\em Geophysical research letters}, 33(16), 2006.

\bibitem{hendon2007seasonal}
Harry~H Hendon, Matthew~C Wheeler, and Chidong Zhang.
\newblock {Seasonal dependence of the {MJO}--{ENSO} relationship}.
\newblock {\em Journal of climate}, 20(3):531--543, 2007.

\bibitem{wang2024insights}
Jinyu Wang, Xianghui Fang, Nan Chen, and Mu~Mu.
\newblock Insights of dynamic forcing effects of {MJO} on {ENSO} from a {Shallow Water Model}.
\newblock {\em Journal of Climate}, 37(23):6143--6166, 2024.

\bibitem{moon2011enso}
Ja-Yeon Moon, Bin Wang, and Kyung-Ja Ha.
\newblock {ENSO} regulation of {MJO} teleconnection.
\newblock {\em Climate dynamics}, 37:1133--1149, 2011.

\bibitem{lee2019enso}
Robert~W Lee, Steven~J Woolnough, Andrew~J Charlton-Perez, and Frederic Vitart.
\newblock {ENSO} modulation of {MJO} teleconnections to the north atlantic and europe.
\newblock {\em Geophysical Research Letters}, 46(22):13535--13545, 2019.

\bibitem{jin1997equatorial}
Fei-Fei Jin.
\newblock {An equatorial ocean recharge paradigm for {ENSO}. Part I: Conceptual model}.
\newblock {\em Journal of the atmospheric sciences}, 54(7):811--829, 1997.

\bibitem{suarez1988delayed}
Max~J Suarez and Paul~S Schopf.
\newblock {A delayed action oscillator for ENSO}.
\newblock {\em Journal of Atmospheric Sciences}, 45(21):3283--3287, 1988.

\bibitem{battisti1989interannual}
David~S Battisti and Anthony~C Hirst.
\newblock {Interannual variability in a tropical atmosphere--ocean model: Influence of the basic state, ocean geometry and nonlinearity}.
\newblock {\em Journal of the Atmospheric Sciences}, 46(12):1687--1712, 1989.

\bibitem{ren2013recharge}
Hong-Li Ren and Fei-Fei Jin.
\newblock {Recharge oscillator mechanisms in two types of ENSO}.
\newblock {\em Journal of Climate}, 26(17):6506--6523, 2013.

\bibitem{weisberg1997western}
Robert~H Weisberg and Chunzai Wang.
\newblock {A western {P}acific oscillator paradigm for the {E}l {N}i{\~n}o-{S}outhern {O}scillation}.
\newblock {\em Geophysical Research Letters}, 24(7):779--782, 1997.

\bibitem{zebiak1987model}
Stephen~E Zebiak and Mark~A Cane.
\newblock {A model El Ni{\~n}o--Southern Oscillation}.
\newblock {\em Monthly Weather Review}, 115(10):2262--2278, 1987.

\bibitem{chen2017enso}
Chen Chen, Mark~A Cane, Andrew~T Wittenberg, and Dake Chen.
\newblock {ENSO} in the {CMIP5} simulations: Life cycles, diversity, and responses to climate change.
\newblock {\em Journal of Climate}, 30(2):775--801, 2017.

\bibitem{geng2022enso}
Licheng Geng and Fei-Fei Jin.
\newblock {ENSO diversity simulated in a revised Cane-Zebiak model}.
\newblock {\em Frontiers in Earth Science}, 10:899323, 2022.

\bibitem{geng2020two}
Tao Geng, Wenju Cai, and Lixin Wu.
\newblock {Two types of ENSO varying in tandem facilitated by nonlinear atmospheric convection}.
\newblock {\em Geophysical Research Letters}, 47(15):e2020GL088784, 2020.

\bibitem{chen2022multiscale}
Nan Chen, Xianghui Fang, and Jin-Yi Yu.
\newblock A multiscale model for {El Ni{\~n}o} complexity.
\newblock {\em npj Climate and Atmospheric Science}, 5(1):16, 2022.

\bibitem{chen2023simple}
Nan Chen and Xianghui Fang.
\newblock {A simple multiscale intermediate coupled stochastic model for {El Ni{\~n}o} diversity and complexity}.
\newblock {\em Journal of Advances in Modeling Earth Systems}, 15(4):e2022MS003469, 2023.

\bibitem{dieppois2021enso}
Bastien Dieppois, Antonietta Capotondi, Benjamin Pohl, Kwok~Pan Chun, Paul-Arthur Monerie, and Jonathan Eden.
\newblock {ENSO} diversity shows robust decadal variations that must be captured for accurate future projections.
\newblock {\em Communications Earth \& Environment}, 2(1):1--13, 2021.

\bibitem{capotondi2013enso}
Antonietta Capotondi.
\newblock {ENSO} diversity in the {NCAR} {CCSM4} climate model.
\newblock {\em Journal of Geophysical Research: Oceans}, 118(10):4755--4770, 2013.

\bibitem{atwood2017characterizing}
Alyssa~R Atwood, David~S Battisti, Andrew~T Wittenberg, WHG Roberts, and Daniel~J Vimont.
\newblock Characterizing unforced multi-decadal variability of {ENSO}: A case study with the {GFDL} {CM}2.1 coupled {GCM}.
\newblock {\em Climate Dynamics}, 49:2845--2862, 2017.

\bibitem{jin2007ensemble}
Fei-Fei Jin, L~Lin, A~Timmermann, and J~Zhao.
\newblock {Ensemble-mean dynamics of the {ENSO} recharge oscillator under state-dependent stochastic forcing}.
\newblock {\em Geophysical research letters}, 34(3), 2007.

\bibitem{vialard2025nino}
J~Vialard, F-F Jin, MJ~McPhaden, A~Fedorov, W~Cai, S-I An, D~Dommenget, X~Fang, MF~Stuecker, C~Wang, et~al.
\newblock {The El Ni{\~n}o Southern Oscillation (ENSO) recharge oscillator conceptual model: Achievements and future prospects}.
\newblock {\em Reviews of Geophysics}, 63(1):e2024RG000843, 2025.

\bibitem{miura2007madden}
Hiroaki Miura, Masaki Satoh, Tomoe Nasuno, Akira~T Noda, and Kazuyoshi Oouchi.
\newblock A {Madden-Julian Oscillation} event realistically simulated by a global cloud-resolving model.
\newblock {\em Science}, 318(5857):1763--1765, 2007.

\bibitem{miyakawa2014madden}
Tomoki Miyakawa, Masaki Satoh, Hiroaki Miura, Hirofumi Tomita, Hisashi Yashiro, Akira~T Noda, Yohei Yamada, Chihiro Kodama, Masahide Kimoto, and Kunio Yoneyama.
\newblock {Madden--Julian Oscillation} prediction skill of a new-generation global model demonstrated using a supercomputer.
\newblock {\em Nature communications}, 5(1):3769, 2014.

\bibitem{lin2024assessment}
Qiao-Jun Lin, V{\'\i}ctor~C Mayta, and {\'A}ngel~F Adames~Corraliza.
\newblock {Assessment of the Madden-Julian Oscillation in CMIP6 models based on moisture mode theory}.
\newblock {\em Geophysical Research Letters}, 51(8):e2023GL106693, 2024.

\bibitem{le2021underestimated}
Phong~VV Le, Cl{\'e}ment Guilloteau, Antonios Mamalakis, and Efi Foufoula-Georgiou.
\newblock {Underestimated {MJO} variability in CMIP6 models}.
\newblock {\em Geophysical research letters}, 48(12):e2020GL092244, 2021.

\bibitem{majda2009skeleton}
Andrew~J Majda and Samuel~N Stechmann.
\newblock {The skeleton of tropical intraseasonal oscillations}.
\newblock {\em Proceedings of the National Academy of Sciences}, 106(21):8417--8422, 2009.

\bibitem{adames2016mjo}
{\'A}ngel~F Adames and Daehyun Kim.
\newblock {The {MJO} as a dispersive, convectively coupled moisture wave: Theory and observations}.
\newblock {\em Journal of the Atmospheric Sciences}, 73(3):913--941, 2016.

\bibitem{yang2013triggered}
Da~Yang and Andrew~P Ingersoll.
\newblock Triggered convection, gravity waves, and the {MJO}: {A} shallow-water model.
\newblock {\em Journal of the atmospheric sciences}, 70(8):2476--2486, 2013.

\bibitem{wang2016trio}
Bin Wang, Fei Liu, and Guosen Chen.
\newblock {A trio-interaction theory for {Madden--Julian} Oscillation}.
\newblock {\em Geoscience Letters}, 3:1--16, 2016.

\bibitem{jiang2020fifty}
Xianan Jiang, {\'A}ngel~F Adames, Daehyun Kim, Eric~D Maloney, Hai Lin, Hyemi Kim, Chidong Zhang, Charlotte~A DeMott, and Nicholas~P Klingaman.
\newblock Fifty years of research on the {Madden-Julian Oscillation}: {R}ecent progress, challenges, and perspectives.
\newblock {\em Journal of Geophysical Research: Atmospheres}, 125(17):e2019JD030911, 2020.

\bibitem{thual2018tropical}
Sulian Thual, Andrew~J Majda, and Nan Chen.
\newblock {A tropical stochastic skeleton model for the MJO, El Ni{\~n}o, and dynamic Walker circulation: A simplified GCM}.
\newblock {\em Journal of Climate}, 31(22):9261--9282, 2018.

\bibitem{moser2024stochastic}
Charlotte Moser, Nan Chen, and Yinling Zhang.
\newblock {A Stochastic Conceptual Model for the Coupled {ENSO} and {MJO}}.
\newblock {\em arXiv preprint arXiv:2411.05264}, 2024.

\bibitem{yang2021enso}
Qiu Yang, Andrew~J Majda, and Nan Chen.
\newblock {{ENSO} diversity in a tropical stochastic skeleton model for the {MJO}, {El Ni{\~n}o}, and dynamic Walker circulation}.
\newblock {\em Journal of Climate}, 34(9):3481--3501, 2021.

\bibitem{marshall2009coupled}
AG~Marshall, O~Alves, and HH~Hendon.
\newblock A coupled {GCM} analysis of {MJO} activity at the onset of {El Ni{\~n}o}.
\newblock {\em Journal of the atmospheric sciences}, 66(4):966--983, 2009.

\bibitem{newman2009important}
Matthew Newman, Prashant~D Sardeshmukh, and C{\'e}cile Penland.
\newblock How important is air--sea coupling in {ENSO} and {MJO} evolution?
\newblock {\em Journal of Climate}, 22(11):2958--2977, 2009.

\bibitem{capotondi2015optimal}
Antonietta Capotondi and Prashant~D Sardeshmukh.
\newblock {Optimal precursors of different types of {ENSO} events}.
\newblock {\em Geophysical Research Letters}, 42(22):9952--9960, 2015.

\bibitem{behringer2004evaluation}
David Behringer and Yan Xue.
\newblock {Evaluation of the global ocean data assimilation system at NCEP: The Pacific Ocean}.
\newblock In {\em Eighth Symp. on Integrated Observing and Assimilation Systems for Atmosphere, Oceans, and Land Surface}, 2004.

\bibitem{huang2017extended}
Boyin Huang, Peter~W Thorne, Viva~F Banzon, Tim Boyer, Gennady Chepurin, Jay~H Lawrimore, Matthew~J Menne, Thomas~M Smith, Russell~S Vose, and Huai-Min Zhang.
\newblock Extended reconstructed sea surface temperature, version 5 {(ERSSTv5)}: upgrades, validations, and intercomparisons.
\newblock {\em Journal of Climate}, 30(20):8179--8205, 2017.

\bibitem{liebmann1996description}
Brant Liebmann and Catherine~A Smith.
\newblock Description of a complete (interpolated) outgoing longwave radiation dataset.
\newblock {\em Bulletin of the American Meteorological Society}, 77(6):1275--1277, 1996.

\bibitem{stechmann2015identifying}
Samuel~N Stechmann and Andrew~J Majda.
\newblock Identifying the skeleton of the {Madden--Julian Oscillation} in observational data.
\newblock {\em Monthly Weather Review}, 143(1):395--416, 2015.

\bibitem{kug2009two}
Jong-Seong Kug, Fei-Fei Jin, and Soon-Il An.
\newblock {Two types of El Ni{\~n}o events: cold tongue {E}l {N}i{\~n}o and warm pool {E}l {N}i{\~n}o}.
\newblock {\em Journal of Climate}, 22(6):1499--1515, 2009.

\bibitem{wang2019historical}
Bin Wang, Xiao Luo, Young-Min Yang, Weiyi Sun, Mark~A Cane, Wenju Cai, Sang-Wook Yeh, and Jian Liu.
\newblock {Historical change of {E}l {N}i{\~n}o properties sheds light on future changes of extreme {E}l {N}i{\~n}o}.
\newblock {\em Proceedings of the National Academy of Sciences}, 116(45):22512--22517, 2019.

\bibitem{chiodi2014subseasonal}
Andrew~M Chiodi, Don~E Harrison, and Gabriel~A Vecchi.
\newblock {Subseasonal atmospheric variability and {El Ni{\~n}o} waveguide warming: Observed effects of the {Madden--Julian Oscillation} and westerly wind events}.
\newblock {\em Journal of Climate}, 27(10):3619--3642, 2014.

\bibitem{gill1980some}
Adrian~E Gill.
\newblock Some simple solutions for heat-induced tropical circulation.
\newblock {\em Quarterly Journal of the Royal Meteorological Society}, 106(449):447--462, 1980.

\bibitem{matsuno1966quasi}
Taroh Matsuno.
\newblock Quasi-geostrophic motions in the equatorial area.
\newblock {\em Journal of the Meteorological Society of Japan. Ser. II}, 44(1):25--43, 1966.

\bibitem{vallis2016geophysical}
Geoffrey~K Vallis.
\newblock {Geophysical fluid dynamics: whence, whither and why?}
\newblock {\em Proceedings of the Royal Society A: Mathematical, Physical and Engineering Sciences}, 472(2192):20160140, 2016.

\bibitem{thual2014stochastic}
Sulian Thual, Andrew~J Majda, and Samuel~N Stechmann.
\newblock A stochastic skeleton model for the {MJO}.
\newblock {\em Journal of the Atmospheric Sciences}, 71(2):697--715, 2014.

\bibitem{thual2016simple}
Sulian Thual, Andrew~J Majda, Nan Chen, and Samuel~N Stechmann.
\newblock Simple stochastic model for {El Ni{\~n}o} with westerly wind bursts.
\newblock {\em Proceedings of the National Academy of Sciences}, 113(37):10245--10250, 2016.

\bibitem{zhao2021breakdown}
Sen Zhao, Fei-Fei Jin, Xiaoyu Long, and Mark~A Cane.
\newblock On the breakdown of {ENSO}'s relationship with thermocline depth in the central-equatorial {Pacific}.
\newblock {\em Geophysical Research Letters}, 48(9):e2020GL092335, 2021.

\bibitem{chen2016filtering}
Nan Chen and Andrew~J Majda.
\newblock {Filtering the stochastic skeleton model for the {Madden--Julian} Oscillation}.
\newblock {\em Monthly Weather Review}, 144(2):501--527, 2016.

\bibitem{stachnik2015evaluating}
Justin~P Stachnik, Duane~E Waliser, Andrew~J Majda, Samuel~N Stechmann, and Sulian Thual.
\newblock Evaluating {MJO} event initiation and decay in the skeleton model using an {RMM}-like index.
\newblock {\em Journal of Geophysical Research: Atmospheres}, 120(22):11--486, 2015.

\bibitem{ogrosky2015mjo}
H~Reed Ogrosky and Samuel~N Stechmann.
\newblock The {MJO} skeleton model with observation-based background state and forcing.
\newblock {\em Quarterly Journal of the Royal Meteorological Society}, 141(692):2654--2669, 2015.

\bibitem{harrison1997westerly}
DE~Harrison and Gabriel~A Vecchi.
\newblock {Westerly wind events in the tropical Pacific, 1986--95}.
\newblock {\em Journal of climate}, 10(12):3131--3156, 1997.

\bibitem{tziperman2007quantifying}
Eli Tziperman and Lisan Yu.
\newblock {Quantifying the dependence of westerly wind bursts on the large-scale tropical Pacific SST}.
\newblock {\em Journal of climate}, 20(12):2760--2768, 2007.

\bibitem{vecchi2000tropical}
Gabriel~A Vecchi and DE~Harrison.
\newblock {Tropical Pacific sea surface temperature anomalies, {El Ni{\~n}o}, and equatorial westerly wind events}.
\newblock {\em Journal of climate}, 13(11):1814--1830, 2000.

\bibitem{hu2016exceptionally}
Shineng Hu and Alexey~V Fedorov.
\newblock {Exceptionally strong easterly wind burst stalling {El Ni{\~n}o} of 2014}.
\newblock {\em Proceedings of the National Academy of Sciences}, 113(8):2005--2010, 2016.

\bibitem{hu2019extreme}
Shineng Hu and Alexey~V Fedorov.
\newblock The extreme {El Ni{\~n}o} of 2015--2016: The role of westerly and easterly wind bursts, and preconditioning by the failed 2014 event.
\newblock {\em Climate Dynamics}, 52(12):7339--7357, 2019.

\bibitem{bergman2001intraseasonal}
John~W Bergman, Harry~H Hendon, and Klaus~M Weickmann.
\newblock Intraseasonal air--sea interactions at the onset of {El Ni{\~n}o}.
\newblock {\em Journal of Climate}, 14(8):1702--1719, 2001.

\bibitem{kapur2012multiplicative}
Atul Kapur and Chidong Zhang.
\newblock Multiplicative {MJO} forcing of {ENSO}.
\newblock {\em Journal of Climate}, 25(23):8132--8147, 2012.

\bibitem{zhang1995relationship}
Guang~Jun Zhang and Michael~J McPhaden.
\newblock The relationship between sea surface temperature and latent heat flux in the equatorial {Pacific}.
\newblock {\em Journal of climate}, 8(3):589--605, 1995.

\bibitem{liu2020impact}
Jia Liu, Yuqin Da, Tim Li, and Feng Hu.
\newblock Impact of {ENSO} on {MJO} pattern evolution over the maritime continent.
\newblock {\em Journal of Meteorological Research}, 34(6):1151--1166, 2020.

\bibitem{dasgupta2021interannual}
Panini Dasgupta, MK~Roxy, Rajib Chattopadhyay, CV~Naidu, and Abirlal Metya.
\newblock Interannual variability of the frequency of {MJO} phases and its association with two types of {ENSO}.
\newblock {\em Scientific reports}, 11(1):11541, 2021.

\bibitem{yu2012identifying}
Jin-Yi Yu and Seon~Tae Kim.
\newblock Identifying the types of major {E}l {N}i{\~n}o events since 1870.
\newblock {\em International journal of climatology}, 33(8):2105--2112, 2012.

\bibitem{chen2015strong}
Dake Chen, Tao Lian, Congbin Fu, Mark~A Cane, Youmin Tang, Raghu Murtugudde, Xunshu Song, Qiaoyan Wu, and Lei Zhou.
\newblock Strong influence of westerly wind bursts on {E}l {N}i{\~n}o diversity.
\newblock {\em Nature Geoscience}, 8(5):339--345, 2015.

\bibitem{freund2019higher}
Mandy~B Freund, Benjamin~J Henley, David~J Karoly, Helen~V McGregor, Nerilie~J Abram, and Dietmar Dommenget.
\newblock Higher frequency of central {P}acific {E}l {N}i{\~n}o events in recent decades relative to past centuries.
\newblock {\em Nature Geoscience}, 12(6):450--455, 2019.

\bibitem{an2000interdecadal}
Soon-Il An and Bin Wang.
\newblock Interdecadal change of the structure of the {ENSO} mode and its impact on the {ENSO} frequency.
\newblock {\em Journal of Climate}, 13(12):2044--2055, 2000.

\bibitem{mcphaden2011nino}
MJ~McPhaden, T~Lee, and D~McClurg.
\newblock {El Ni{\~n}o} and its relationship to changing background conditions in the tropical {Pacific Ocean}.
\newblock {\em Geophysical Research Letters}, 38(15), 2011.

\bibitem{xiang2013new}
Baoqiang Xiang, Bin Wang, and Tim Li.
\newblock A new paradigm for the predominance of standing central {Pacific} warming after the late 1990s.
\newblock {\em Climate Dynamics}, 41:327--340, 2013.

\bibitem{power2021decadal}
Scott Power, Matthieu Lengaigne, Antonietta Capotondi, Myriam Khodri, J{\'e}r{\^o}me Vialard, Beyrem Jebri, Eric Guilyardi, Shayne McGregor, Jong-Seong Kug, Matthew Newman, et~al.
\newblock Decadal climate variability in the tropical {P}acific: {C}haracteristics, causes, predictability, and prospects.
\newblock {\em Science}, 374(6563):eaay9165, 2021.

\bibitem{chen2023stochastic}
Nan Chen.
\newblock {\em Stochastic methods for modeling and predicting complex dynamical systems}.
\newblock Springer, 2023.

\bibitem{stein2010seasonal}
Karl Stein, Niklas Schneider, Axel Timmermann, and Fei-Fei Jin.
\newblock Seasonal synchronization of {ENSO} events in a linear stochastic model.
\newblock {\em Journal of Climate}, 23(21):5629--5643, 2010.

\bibitem{tziperman1997mechanisms}
Eli Tziperman, Stephen~E Zebiak, and Mark~A Cane.
\newblock Mechanisms of seasonal--{ENSO} interaction.
\newblock {\em Journal of the Atmospheric Sciences}, 54(1):61--71, 1997.

\bibitem{thual2017seasonal}
Sulian Thual, Andrew Majda, and Nan Chen.
\newblock Seasonal synchronization of a simple stochastic dynamical model capturing {El Ni{\~n}o} diversity.
\newblock {\em Journal of Climate}, 30(24):10047--10066, 2017.

\bibitem{mitchell1992annual}
Todd~P Mitchell and John~M Wallace.
\newblock The annual cycle in equatorial convection and sea surface temperature.
\newblock {\em Journal of Climate}, 5(10):1140--1156, 1992.

\bibitem{seiki2007westerly}
Ayako Seiki and Yukari~N Takayabu.
\newblock Westerly wind bursts and their relationship with intraseasonal variations and {ENSO}. {Part I: Statistics}.
\newblock {\em Monthly Weather Review}, 135(10):3325--3345, 2007.

\bibitem{wheeler1999convectively}
Matthew Wheeler and George~N Kiladis.
\newblock Convectively coupled equatorial waves: {A}nalysis of clouds and temperature in the wavenumber--frequency domain.
\newblock {\em Journal of the Atmospheric Sciences}, 56(3):374--399, 1999.

\bibitem{jauregui2024mjo}
Yakelyn~R Jauregui and Shuyi~S Chen.
\newblock {MJO-Induced Warm Pool Eastward Extension Prior to the Onset of El Ni{\~n}o: Observations from 1998 to 2019}.
\newblock {\em Journal of Climate}, 37(3):855--873, 2024.

\bibitem{kessler2001eof}
William~S Kessler.
\newblock {EOF representations of the {Madden--Julian} Oscillation and its connection with {ENSO}}.
\newblock {\em Journal of Climate}, 14(13):3055--3061, 2001.

\bibitem{fang2023quantifying}
Xianghui Fang and Nan Chen.
\newblock Quantifying the predictability of {ENSO} complexity using a statistically accurate multiscale stochastic model and information theory.
\newblock {\em Journal of Climate}, 36(8):2681--2702, 2023.

\bibitem{wang2025dual}
Jinyu Wang, Xianghui Fang, Nan Chen, Bo~Qin, Mu~Mu, and Chaopeng Ji.
\newblock A dual-core model for {ENSO} diversity: Unifying model hierarchies for realistic simulations.
\newblock {\em npj Clim Atmos Sci}, 8(269):5285--5310, 2025.

\bibitem{zhang2025data}
Yinling Zhang.
\newblock {Dataset for Publication ``A Simple Intermediate Coupled MJO-ENSO Model: Multiscale Interactions and ENSO Complexity''}.
\newblock \url{https://doi.org/10.5281/zenodo.16113371}, 2025.
\newblock [Data set].

\bibitem{zhang2025code}
Yinling Zhang.
\newblock {Codes for Publication ``A Simple Intermediate Coupled MJO-ENSO Model: Multiscale Interactions and ENSO Complexity''}.
\newblock \url{https://doi.org/10.5281/zenodo.16113134}, 2025.
\newblock [Software].

\bibitem{majda2003introduction}
Andrew Majda.
\newblock {\em Introduction to {PDE}s and Waves for the Atmosphere and Ocean}, volume~9.
\newblock American Mathematical Soc., 2003.

\bibitem{majda2003systematic}
Andrew~J Majda and Rupert Klein.
\newblock Systematic multiscale models for the tropics.
\newblock {\em Journal of the Atmospheric Sciences}, 60(2):393--408, 2003.

\bibitem{stechmann2014walker}
Samuel~N Stechmann and H~Reed Ogrosky.
\newblock The {W}alker circulation, diabatic heating, and outgoing longwave radiation.
\newblock {\em Geophysical Research Letters}, 41(24):9097--9105, 2014.

\end{thebibliography}

\end{document}